\documentclass[twocolumn,nofootinbib]{revtex4}
\usepackage{amsfonts}
\usepackage{amsmath}
\usepackage{amssymb}
\usepackage{xcolor}

\usepackage{tensor}

\newcommand{\pk}{\mbox{$k_{\parallel}$}}

\newcommand{\kp}{\mbox{$k_{\parallel}$}}

\newcommand{\ik}{\frac{ik_{\parallel}}{a_1}}

\definecolor{blue}{rgb}{0,0,1}
\definecolor{green}{rgb}{0,1,0}
\definecolor{turkos}{rgb}{0,0.4,0.4}
\definecolor{red}{rgb}{1,0,0}
\definecolor{darkblue}{rgb}{0,0,0.5}

\newcommand{\rka}{\frac{\mathcal{R}a_2^2-k_{\perp}^2}{2a_2}}

\newcommand{\okt}{\mbox{$k_{\perp}^2$}}
\newcommand{\pkt}{\mbox{$k_{\parallel}^2$}}

\input{tcilatex}
\begin{document}

\title{General Perfect Fluid Perturbations of Homogeneous  and Orthogonal \\
Locally Rotationally Symmetric Class II
Cosmologies
}
\author{Robin T\"ornkvist$^{1}$ and Michael Bradley$^{2}$}
\address{$^{1}$\textit{Department of Astronomy and Theoretical Physics, Lund University, Lund, Sweden},\\
$^{2}$\textit{Department of Physics, Ume{\aa } University, Ume\aa, Sweden,} \\
robin.tornkvist@thep.lu.se, michael.bradley@physics.umu.se
}
\date{\today }

\begin{abstract}
First order perturbations of homogeneous and hypersurface orthogonal LRS (Locally Rotationally Symmetric) class II cosmologies with a cosmological constant are considered in the framework of the 1+1+2 covariant decomposition of spacetime. The perturbations, which are of perfect fluid type, include general scalar, vector and tensor modes  and extend some previous works where vorticity perturbations were excluded.
 A harmonic decomposition is performed and the field equations are then reduced to a set of eight evolution equations for eight harmonic coefficients, representing perturbations in density, shear, vorticity and the Weyl tensor, in terms of which all other variables can be expressed algebraically. This system decouples into two sub-systems, one for five and one for three coefficients. As previously known, vorticity perturbations cannot be generated to any order in a barytopic perfect fluid. Hence the time development of existing first order vorticity perturbations  are seen to be completely determined by the background. However, an already existing vorticity will act as source terms in the evolution equations for the other quantities.
 In the high frequency approximation the four independent Weyl tensor harmonics evolve as gravitational waves on the anisotropic background in the same manner as in the case without vorticity, whereas vorticity gives a first order disturbance of sonic waves.
\end{abstract}

\maketitle

\section{Introduction}

Present observations of cosmological parameters, like the large scale structure, anisotropies
of the background radiation (CMB) and the cosmological redshift are well described by an expanding
isotropic and homogeneous background metric, where the time development is sourced by
a cosmological constant and a fluid consisting of baryonic matter, radiation and dark matter,
the  $\Lambda$CDM model \cite{WMAP9yr,
Planck2}. However, there are deviations. For example, the predicted
amplitudes in the power spectrum of the CMB differ from the observed for large angles \cite{Bennett46,Oliveria45,Vielva59,PlanckAnomaly},
even though the uncertainties are high for these angles,
and the upper limits of a possible anisotropy in the cosmological redshift are rather big \cite{H1,H2,Dec}.
Hence it is of interest to investigate the properties of other models, like, e.g. the evolution of perturbations on anistropic
backgrounds. If the background has one spatial direction of anisotropy, so that it is
Locally Rotationally Symmetric (LRS), a gauge invariant perturbation theory built on the 1+1+2 covariant split of spacetime \cite{1+1+2,Schperturb,LRSIIscalar},
which generalizes the 1+3 covariant split \cite{cov1,Cargese,cov3,cov5,EMM}, is
suitable for the study. For other perturbative methods used in general relatvity see e.g. \cite{Lifshitz,Bardeen,Stewart,Hawking,Olson} and for
some previous works on perturbations in anisotropic universes see e.g. \cite{Doroschkevich,Perko,Tomita,Gumruk,Pereira,Pitrou}.

In some earlier papers \cite{GWKS,BFK,KASAscalar} we studied vorticity free perturbations on a class of 
anisotropic cosmological models, given by the homogeneous and orthogonal locally rotationally symmetric (LRS)
spacetimes of class II \cite{LRS,MarklundBradley}. The restriction to zero vorticity was partly due to the significant simplifications this
implies for the commutator relations, 
but was also motivated by that vorticity cannot be generated in a perfect fluid with barotropic equation
of state, cf. e.g. \cite{Raichoudhuri,Cargese,LuAnandaClarksonMaartens}  and that an existing vorticity in a homogeneous and isotropic universe
decays with time \cite{Hawking}. However, the vorticity acts
as source terms for the other quantities and hence can give imprints on other measurable quantities. For works on vorticity perturbations on Robertson-Walker
backgrounds see
e.g. \cite{Hawking,Christopherson1,ChristophersonMalik,EllisBruniHwang,Novelloetal,LuAnandaClarksonMaartens}.

In this paper we present a general treatment of a class of
perturbed LRS cosmological backgrounds, with nonvanishing vorticity on the perturbed model. 
 We start by assuming a homogeneous and hypersurface orthogonal cosmological background model which is of LRS class II, characterized by that the vorticity, magnetic part of the Weyl tensor, and twist of the 2-sheets all vanish \cite{LRS}. The matter content of the model is assumed to be described by a perfect fluid, and a nonvanishing cosmological constant. The choice of an anisotropic background gives us a preferred spatial direction, which motivates our choice to utilize the 1+1+2 covariant formalism \cite{1+1+2}, in performing the perturbation. This formalism entails first splitting spacetime in the 1+3 covariant formalism \cite{cov1,Cargese}, with the average velocity of matter being the preferred direction for each point. After this split we are left with 3-surfaces in spacetime which we further decompose, this time with the anisotropy as the preferred direction, and this gives us all relevant quantities in the 1+1+2 covariant formalism.

In order for the perturbation to be gauge invariant we define all our variables in terms of quantities that vanish on the background. Due to the Stewart-Walker lemma \cite{StewartWalker} these variables are ensured to be gauge invariant. With the background and perturbation method properly defined, we proceed to linearize the propagation equations and constraints given in the 1+1+2 formalism, and then decompose the gauge invariant variables using a harmonic decomposition. Our initial system, consisting of scalar, vector, and tensor equations, is then reduced to a system containing only scalar time evolution equations and constraints. This system is then solved and the result is a set of eight harmonic coefficients, out of which all other variables can be expressed, and their associated time evolution equations.

The final set of harmonic coefficients decouples into an even and odd sector, containing five respectively three variables. The vorticity appears in both these sectors, with one degree of freedom in each, and its time evolution equations completely decouples from the other variables. Assuming a linear equation of state, we get an analytical solution for the vorticity, which depends on time through the expansion coefficients, and on the comoving wavenumbers through integration constants dependent on the initial conditions. The consistency of this solutions is checked specifically through the fall off rate for the case of dust and radiation, and the conservation of angular momentum.

Even though the evolution of the vorticity is decoupled from the other variables, the vorticity appears as a source term in the time evolution of five of the remaining six harmonic coefficients. We move on to examine these evolution equations in the geometrical optics approximation by deriving second order wavelike equations for the harmonic coefficients in
the limit of large harmonic numbers $k$. On keeping terms to the two highest orders in $k$,  damped shear waves are then found to propagate with the speed of sound and to be sourced by vorticity.
The evolution equations for the four components of the Weyl tensor are to this order unaffected by the vorticity and represent gravitational waves.

The paper is organized as follows: In section \ref{1+3+1+1+2} the theory behind the 1+3 and 1+1+2 covariant splits of spacetime are briefly summarized and in section \ref{sectionbackground}
the cosmological backgrounds are described. The perturbative method is described in section \ref{sectionperturbations} and harmonic decompositions of the first order quantities are shown.
In section \ref{EvolutionEquations} the two reduced systems of evolution equations are given and in section \ref{Vorticity} the equations for the vorticity perturbations
are solved and discussed. Finally, in section \ref{Optics} the high frequency approximation is considered and then the results are summarized in section \ref{conclusions}. The commutation relations between the different differential operators are given in appendix A
and properties of the harmonics in appendix B. The harmonic expansion of the linearized equations are presented in appendix C and then 27 harmonic coefficients are solved for algebraically in terms of the
8 remaining in appendix D.

\section{The 1 + 3 and 1 + 1 + 2 Covariant Splits of Spacetime}\label{1+3+1+1+2}

\label{sectioncovariant}

We will in this paper utilize the 1 + 1 + 2 covariant split of spacetime, since we have two preferred directions on the background. This method builds on the 1 + 3 covariant split of spacetime, and in this section we will provide a brief summary of both these formalisms. More details about the 1 + 3 formalism can be found in \cite{Cargese, cov1}, and about the 1 + 1 + 2 formalism in \cite{1+1+2, Schperturb}.

We start by looking at the 1 + 3 covariant split of spacetime, which is a suitable method when there exists a preferred unit timelike vector $u^a$. We use this vector $u^a$ to define a projection tensor $\tensor{h}{_a^b} \equiv \tensor{g}{_a^b}+u_au^b$, where $g_{ab}$ is the 4-metric, which projects onto the 3-surfaces orthogonal to $u^a$. This projection tensor allows us to covariantly decompose vectors and tensors into spatial and timelike parts. Using $u^a$ and $h_{ab}$ we can define the covariant time derivative and projected spatial derivative as

\begin{equation}
\tensor{\dot{T}}{_{a\cdots b}^{c\cdots d}} \equiv u^e\nabla_e\tensor{T}{_{a\cdots b}^{c\cdots d}}
\end{equation}
and
\begin{equation}
D_e\tensor{T}{_{a\cdots b}^{c\cdots d}} \equiv \tensor{h}{_a^f}\cdots\tensor{h}{_b^g}\tensor{h}{_p^c}\cdots\tensor{h}{_q^d}\tensor{h}{^r_e}\nabla_r\tensor{T}{_{f\cdots g}^{p\cdots q}},
\end{equation}

\noindent respectively.

The independent variables of relevance are now given by the decomposed form of the Ricci tensor, the Weyl tensor and the covariant derivative of $u^a$. The Ricci tensor can be expressed, using Einstein's field equations, through the energy momentum tensor which is decomposed as $T_{ab}=\mu u_au_b+ph_{ab}$ for a perfect fluid. Here, $\mu \equiv T_{ab}u^au^b$ is the energy density and $p \equiv \frac{1}{3}T_{ab}h^{ab}$ is the isotropic pressure of the fluid. The Weyl tensor is decomposed into an electric and magnetic part, $E_{ab} \equiv C_{acbd}u^cu^d$ and $H_{ab} \equiv \frac{1}{2}\varepsilon_{ade}\tensor{C}{^d^e_b_c}u^c$, respectively, where we have introduced the volume element on the 3-surfaces $\varepsilon_{abc}\equiv \eta _{dabc}u^{d}\equiv 4!\sqrt{-g}\delta _{\lbrack d}^{0}\delta _{a}^{1}\delta _{b}^{2}\delta _{c]}^{3}u^{d}$. Lastly, the covariant derivative of $u^a$ is decomposed as

\begin{equation}
	\nabla_au_b=-u_aA_b+\frac{1}{3}\Theta h_{ab}+\sigma_{ab}+\omega_{ab}
\end{equation}

\noindent where we have the acceleration $A_a \equiv u^b\nabla_b u_a$, rate of volume expansion $\Theta \equiv D_a u^a$, rate of shear $\sigma_{ab} \equiv D_{\langle a}u_{b \rangle}$ and vorticity $\omega_{ab} \equiv D_{[a}u_{b]}$. Here the angular brackets denotes the orthogonally projected trace-free symmetric part of a tensor, defined as $T_{\langle ab \rangle} \equiv \left(\tensor{h}{_{(a}^c}\tensor{h}{_{b)}^d}-\frac{1}{3}h_{ab}h^{cd} \right)T_{cd}$, and the square brackets denotes the antisymmetric part of a tensor. We also further define a vorticity vector as $\omega^a \equiv \frac{1}{2}\varepsilon^{abc}\omega_{bc}$. The evolution in the direction of $u^a$ and the constraints for the independent variables introduced above are now given by the Ricci identity for $u^a$, the once and twice contracted Bianchi identities and the commutation relations of the new projected derivatives, and can be found in e.g. \cite{Cargese}.

We can further decompose our physical quantities in what is known as the 1+1+2 covariant formalism. This uses a preferred spatial direction, given in terms of a unit vector $n^a$ which is orthogonal to $u^a$, i.e. it resides on the 3-surfaces. As previously, we introduce a projection tensor $\tensor{N}{_a^b}=\tensor{h}{_a^b}-n_an^b$ which will project vectors and tensors onto the 2-sheets orthogonal to $n^a$. This allows us to decompose any spatial vector into a scalar part along $n^a$ and a vector part on the 2-sheet, and any spatial tensor into a scalar part along $n^a$, a vector part on the 2-sheet and a projected, trace-free and symmetric 2-tensor. We also have the derivative along $n^a$ and the projected derivative on the 2-surfaces defined as

\begin{equation}
  \tensor{\widehat{T}}{_{a\cdots b}^{c\cdots d}} = n^eD_e\tensor{T}{_{a\cdots b}^{c\cdots d}}
\end{equation}

\noindent and

\begin{equation}
  \delta_e\tensor{T}{_{a\cdots b}^{c\cdots d}}=\tensor{N}{_e^j}\tensor{N}{_a^f}\cdots \tensor{N}{_b^g}\tensor{N}{_h^c}\cdots \tensor{N}{_i^d}D_j\tensor{T}{_{f\cdots g}^{h\cdots i}},
\end{equation}

\noindent respectively.

Decomposing the kinematic quantities given in the 1 + 3 formalism gives us $A^a =  \mathcal{A}n^a+\mathcal{A}^a$, $\omega^a = \Omega n^a + \Omega^a$ and $\sigma_{ab} = \Sigma\left(n_an_b - \frac{1}{2}N_{ab}\right)+ 2\Sigma_{(a}n_{b)}+\Sigma_{ab}$. The electric and magnetic part of the Weyl tensor, $E_{ab}$ and $H_{ab}$, are decomposed in the same way as the shear, in terms of the variables $\mathcal{E}$, $\mathcal{E}_a$, $\mathcal{E}_{ab}$ and $\mathcal{H}$, $\mathcal{H}_a$, $\mathcal{H}_{ab}$. We also get new kinematic quantities by decomposing $D_an_b$ and $\dot{n}_a$ as

\begin{equation}
	D_a n_b= n_a a_b + \frac{1}{2}\phi N_{ab} + \xi \varepsilon_{ab}+\zeta_{ab} 
\end{equation}
and
\begin{equation}
	 \dot{n}^a = \mathcal{B}u^a+\alpha^a,
\end{equation}

\noindent where we have the acceleration of $n^a$, $a^a \equiv n^cD_cn^a$, the 2-sheet expansion $\phi \equiv \delta_a n^a$, the twisting of the 2-sheets $\xi \equiv \frac{1}{2}\varepsilon^{ab}\delta_an_b$, the shear of $n^a$, $\zeta_{ab} \equiv \delta_{\{ a} n_{b\}}$, and $\alpha_a \equiv \tensor{N}{_a^b}\dot n_b$. It follows, since $u^an_a = 0$, that $\mathcal{B}=\mathcal{A}$. In the above decomposition we have introduced the volume element on the 2-sheets $\varepsilon _{ab}\equiv \varepsilon _{abc}n^{c}$, and also the shorthand notation of curly brackets to denote the projected, trace-free and symmetric part of a 2-tensor. We will also introduce the notation of a bar over a vector index to denote that index to be projected onto the 2-sheets, i.e. $\psi_{\overline{a}}\equiv\tensor{N}{_a^b}\psi_{b}$. 

The Ricci identities  for $u^a$ and $n^a$ and the Bianchi identities can now be 
decomposed into evolution and propagation equations in the direction of $u^a$ respectively $n^a$, and constraint equations. For the full set of these equations, see e.g. \cite{1+1+2}. We also get commutation relations between the operators $\dot{ }$, $\hat{ }$ and $\delta_a$ which can be found in appendix \ref{commutation}.

\section{Background spacetimes}
\label{sectionbackground}

The properties of the background spacetimes are here breifly summarized. For more details the reader is referred to \cite{LRS,MarklundBradley,BFK}.

As backgrounds we will use the class of homogeneous, hypersurface orthogonal and locally rotational symmetric (LRS) perfect fluid solutions to Einstein's equations with vanishing magnetic part of the Weyl tensor. For technical reasons we exclude the hyperbolic and closed Friedmann models, which are different from the others both in form of line element and harmonic decomposition \cite{BFK}. In the classification of \cite{LRS} the used backgrounds belong to LRS class II.

These spacetimes have two prefered directions, given by the 4-velocity of the fluid, $u^a$, and the direction of anistropy, $n^a$. The 2-sheets perpendicular to $n^a$
are maximally symmetric with 2D curvature scalar ${\mathcal{R}}=2{\mathcal{K}}/a_2^2$, where $a_2(t)$ is the radius of curvature and ${\mathcal{K}}=\pm 1$ or 0 for spheres, pseudo-spheres or planes, respectively. 
In terms of the quantities of the 1+1+2 formalism introduced in section~\ref{sectioncovariant}, these spacetimes are characterized by the following nonzero quantities
\begin{equation}
S^{(0)}= \{\mu,p,{\mathcal{E}},\Theta,\Sigma\} \, .
\end{equation}
On assuming a nonzero cosmological constant $\Lambda$, they satisfy the following set of evolution equations
\begin{equation}
\dot{\mu}=-\Theta \left( \mu +p\right) \ ,  \label{continuity}
\end{equation}%
\begin{equation}
\dot{\Theta}=-\frac{\Theta ^{2}}{3}-\frac{3}{2}\Sigma ^{2}-\frac{1}{2}\left(
\mu +3p\right) +\Lambda \ ,  \label{expansiondot}
\end{equation}%
\begin{equation}
\dot{\Sigma}=\frac{2}{3}\left( \mu +\Lambda \right) +\frac{\Sigma ^{2}}{2}%
-\Sigma \Theta -\frac{2}{9}\Theta ^{2}\ ,  \label{Sigmadot}
\end{equation}
and  ${\mathcal{E}}$ is given algebraically as
\begin{equation}
3\mathcal{E}=-2\left( \mu +\Lambda \right) -3\Sigma ^{2}+\frac{2}{3}\Theta
^{2}+\Sigma \Theta \ .  \label{EWeyl}
\end{equation}
Hence a closed system is obtained if an equation of state $p=p(\mu)$ is given. The 2D curvature scalar may be expressed
in terms of the quantities in $S^{(0)}$ as
 \begin{equation}
\mathcal{R}=\frac{2{\mathcal{K}}}{a_{2}^{2}
}=2\left( \mu +\Lambda \right) +%
\frac{3}{2}\Sigma ^{2}-\frac{2\Theta ^{2}}{3}\ .  \label{2TimesGaussianCurv}
\end{equation}

The line-element can for this class of metrics be written as 
\begin{equation}  \label{metric}
ds^{2}=-dt^{2}+a_{1}^{2}\left( t\right) dz^{2}+a_{2}^{2}\left( t\right)
\left( d\vartheta ^{2}+f_{\mathcal{K}}(\vartheta) d\varphi ^{2}\right) \ , 
\end{equation}%
in terms of the two scale factors $a_1(t)$ and $a_2(t)$. For spheres (${\cal{K}}=1$) $f_1(\vartheta)=\sin^2\vartheta$,
for pseudo-spheres (${\cal{K}}=-1$) $f_{-1}(\vartheta)=\sinh^2\vartheta$ and for planes (${\cal{K}}=0$) 
 $f_{0}(\vartheta)=1$  (or $f_0(\vartheta)=\vartheta^2$).  
The expansion and shear are then given by
\begin{equation}
\Theta =\frac{\dot{a}_{1}}{a_{1}}+2\frac{\dot{a}_{2}}{a_{2}}\ ,
\label{Thetabackgroundexp}
\end{equation}%
\begin{equation}
\Sigma =\frac{2}{3}\left( \frac{\dot{a}_{1}}{a_{1}}-\frac{\dot{a}_{2}}{a_{2}}%
\right) \; .  \label{Sigmabackgroundexp}
\end{equation}

\section{Perturbations}
\label{sectionperturbations}

In some previous papers we have studied perturbations on the backgrounds given in section \ref{sectionbackground} \cite{GWKS,BFK}. Here these studies will be extended 
to the most general first order perturbations, consistent with a perfect fluid, by including also vorticity of the fluid. To avoid
the gauge problem in perturbation theory, i.e. the problem of identifying the perturbed spacetime with the background, we will
use an approach based on the covariant split of spacetime and quantities which are zero on the background.

As variabels we will use the scalars which are nonzero on the background, i.e. $G=\{\mu, p, {\mathcal{E}}, \Theta, \Sigma\}$,
and the remaining 1+1+2 covariantly defined quantites from section \ref{sectioncovariant}, which are zero on the background. The latter are gauge-invariant due to the 
Stewart-Walker lemma \cite{StewartWalker}. To get a consistent first order system we need also expand the quantities $G$ as $G=G^{(0)}+G^{(1)}$,
e.g. $\mu=\mu^{(0)}+\mu^{(1)}$ etc. The $G^{(1)}$ can in a convenient way be represented by the gradients of the $G$, i.e. by 
$\delta_a G$ and $\hat G$. These are also zero on the background for which the quantities only are functions of time. 
As will be shown later in section \ref{subsectionrelharmonics}, where a harmonic decomposition of the first order quantities
is done, the $\hat G$ derivatives may be expressed in terms of the derivatives $Z_a\equiv\delta_a G$. Hence, instead of $G^{(1)}$,
we introduce the following first order quantities $Z_a$

\begin{gather}  \label{newvariables}\nonumber
\mu_a=\delta_a\mu \, ,\;\; p_a=\delta_a p \, ,\;\; W_a=\delta_a\Theta, \\
 V_a=\delta_a \Sigma \, ,\;\; X_a=\delta_a{\mathcal{E}} \,
\end{gather}

\noindent which vanish on the background. The complete set of first order quantities which vanish on the background is then given by
\begin{equation}
\begin{aligned}
S^{(1)} & \equiv \{ X_{a},V_{a},W_{a},\mu _{a},p_{a},{\mathcal{A}},{%
\mathcal{A}}_{a},\Sigma _{a},\Sigma _{ab},{\mathcal{E}}_{a},{\mathcal{E}}%
_{ab}, \\
& \quad \quad {\mathcal{H}},{\mathcal{H}}_{a},{\mathcal{H}}_{ab},a_{b},\alpha_a,\phi
,\xi ,\zeta _{ab},\Omega,\Omega_a\} ~.
\end{aligned}
\end{equation}
The frame is partly locked by using the 4-velocity of the fluid as the preferred timelike vector $u^a$ also for the perturbed spacetime.
The direction $n^a$ of anistropy is not well-defined in the perturbed spacetime and we will later restrict the frame, see section \ref{EvolutionEquations}. 

On imposing the Ricci identities for the vectors $u^a$ and $n^a$ and the Bianchi identities together with the commutator relations (see appendix \ref{commutation}),
evolution equations along $u^a$, propagation equations along $n^a$ and constraints are obtained for the covariant quantities. The exact 
equations are found in \cite{1+1+2}. The linearized equations for the first order quantities are given below.

\subsection{Linearized Equations}\label{sectionLE}

The evolution equations are

\begin{equation}
    \dot{\phi} = \left(\Sigma-\frac{2\Theta}{3} \right)\left(\frac{1}{2}\phi-\mathcal{A} \right)+\delta_a \alpha^a,
\end{equation}

\begin{equation}
    \dot{\xi} = \frac{1}{2}\left(\Sigma-\frac{2\Theta}{3} \right) \xi+\frac{1}{2}\varepsilon_{ab}\delta^a\alpha^b+\frac{1}{2}\mathcal{H},
\end{equation}

\begin{equation}
    \dot{\zeta}_{\{ab\}} = \frac{1}{2}\left(\Sigma-\frac{2\Theta}{3} \right)\zeta_{ab}+\delta_{\{a}\alpha_{b\}}-\varepsilon_{c\{a}\tensor{\mathcal{H}}{_{b\}}^c},
\end{equation}

\begin{equation}
    \dot{\Omega} = \left(\Sigma-\frac{2\Theta}{3} \right)\Omega+\frac{1}{2}\varepsilon_{ab}\delta^a\mathcal{A}^b,
\end{equation}

\begin{equation}
    \dot{\Sigma}_{\{ab\}} = \left(\Sigma-\frac{2\Theta}{3} \right)\Sigma_{ab}+\delta_{\{a}\mathcal{A}_{b\}}-\mathcal{E}_{ab},
\end{equation}

\begin{equation}
    \dot{\mathcal{H}} =\frac{3}{2}\left(\Sigma-\frac{2\Theta}{3} \right)\mathcal{H} -\varepsilon_{ab}\delta^a\mathcal{E}^b -3\mathcal{E}\xi,
\end{equation}

\begin{equation}
  \begin{aligned}
    \dot{\mu}_{\overline{a}} &  = \frac{1}{2}\left(\Sigma-\frac{2\Theta}{3} \right)\mu_a+\dot{\mu}\mathcal{A}_a \\
    & \quad -(\mu+p)W_a-\Theta\left( \mu_a+p_a\right),
    \end{aligned}
\end{equation}

\begin{equation}
	\begin{aligned}
        \dot{X}_{\overline{a}} & = \left(\Sigma-\frac{2\Theta}{3} \right)X_a + \dot{\mathcal{E}}\mathcal{A}_a -\frac{1}{2}(\mu+p-3\mathcal{E})V_a \\ 
        & \quad -\frac{1}{2}(\mu_a+p_a)\Sigma -\mathcal{E}W_a+\varepsilon_{bc}\delta_a\delta^b\mathcal{H}^c, \\   
    \end{aligned}
\end{equation}

\begin{eqnarray}\nonumber
        \dot{V}_{\overline{a}}-\frac{2}{3}\dot{W}_{\overline{a}} & =& \frac{3}{2}\left(\Sigma-\frac{2\Theta}{3} \right)\left(V_a-\frac{2}{3}W_a \right) +\frac{1}{3}(\mu_a+3p_a)\\
        && \quad -\delta_a\delta_b\mathcal{A}^b +\left( \dot{\Sigma}-\frac{2\dot{\Theta}}{3}\right)\mathcal{A}_a-X_a  \; .
\end{eqnarray}

\noindent The equations containing a mixture of evolution and propagation contributions are

\begin{equation}
    \widehat{\alpha}_{\overline{a}}-\dot{a}_{\overline{a}} = \left( \Sigma + \frac{\Theta}{3}\right) \mathcal{A}_a+\frac{\Theta}{3} a_a+\Sigma a_A  -\varepsilon_{ab}\mathcal{H}^b,
\end{equation}

\begin{equation}
    \widehat{\mathcal{A}}-\dot{\Theta} = -\delta_a\mathcal{A}^a+\frac{\Theta^2}{3}+\frac{3\Sigma^2}{2}+\frac{1}{2}(\mu+3p)-\Lambda,
\end{equation}

\begin{equation}
    \dot{\Omega}_{\overline{a}}+\frac{1}{2}\varepsilon_{ab}\widehat{\mathcal{A}}^b = -\left( \frac{\Sigma}{2} + \frac{2\Theta}{3} \right)\Omega_a +\frac{1}{2}\varepsilon_{ab}\delta^b\mathcal{A},
\end{equation}

\begin{equation}
    \dot{\Sigma}_{\overline{a}}-\frac{1}{2}\widehat{\mathcal{A}}_{\overline{a}} = -\left( \frac{\Sigma}{2} + \frac{2\Theta}{3} \right)\Sigma_a + \frac{1}{2}\delta_a\mathcal{A} - \frac{3\Sigma}{2}\alpha_a-\mathcal{E}_a,
\end{equation}

\begin{equation}
	\begin{aligned}
        \dot{\mathcal{E}}_{\overline{a}}+\frac{1}{2}\varepsilon_{ab}\widehat{\mathcal{H}}^b  & = \frac{3}{4}\varepsilon_{ab}\delta^b\mathcal{H} - \frac{1}{2}\left(\mu+p-\frac{3\mathcal{E}}{2}\right)\Sigma_a \\
        & \quad +\left(\frac{3\Sigma}{4}-\Theta \right)\mathcal{E}_a+\frac{3\mathcal{E}}{4}\varepsilon_{ab}\Omega^b \\
        & \quad +\frac{1}{2}\varepsilon_{bc}\delta^b\tensor{\mathcal{H}}{^c_a} -\frac{3\mathcal{E}}{2}\alpha_a ,
    \end{aligned}
\end{equation}

\begin{equation}
  \begin{aligned}
    \dot{\mathcal{E}}_{\{ab\}}-\varepsilon_{c\{a}\tensor{\widehat{\mathcal{H}}}{_{b\}}^c} &  = -\varepsilon_{c\{a}\delta^c\mathcal{H}_{b\}} -\frac{1}{2}(\mu+p)\Sigma_{ab} \\
    & \quad -\left( \frac{3\Sigma}{2} + \Theta\right) \mathcal{E}_{ab}-\frac{3\mathcal{E}}{2}\Sigma_{ab},
    \end{aligned}
\end{equation}

\begin{equation}
  \begin{aligned}
    \dot{\mathcal{H}}_{\overline{a}}-\frac{1}{2}\varepsilon_{ab}\widehat{\mathcal{E}}^b &  = -\frac{1}{2}\varepsilon_{bc}\delta^b\tensor{\mathcal{E}}{^c_a}+\left( \frac{3\Sigma}{4}-\Theta\right) \mathcal{H}_a \\
    & \quad +\frac{3\mathcal{E}}{4}\varepsilon_{ab}a^b-\frac{3\mathcal{E}}{2}\varepsilon_{ab}\mathcal{A}^b  -\frac{3}{4}\varepsilon_{ab}X^b,
    \end{aligned}
\end{equation}

\begin{equation}
  \begin{aligned}
    \dot{\mathcal{H}}_{\{ab\}}+\varepsilon_{c\{a}\tensor{\widehat{\mathcal{E}}}{_{b\}}^c} & = \varepsilon_{c\{a}\delta^c\mathcal{E}_{b\}}+\frac{3\mathcal{E}}{2}\varepsilon_{c\{a}\tensor{\zeta}{_{b\}}^c} \\
    & \quad -\left( \frac{3\Sigma}{2} + \Theta\right)\mathcal{H}_{ab} \; .
    \end{aligned}
\end{equation}

\noindent The equations containing only propagation contributions are

\begin{equation}
    \widehat{\phi} = \frac{2\Theta^2}{9}+\frac{\Theta\Sigma}{3} +\delta_a a^a -\frac{2}{3}(\mu+\Lambda)-\mathcal{E}-\Sigma^2,
\end{equation}

\begin{equation}
    \widehat{\xi} = \left( \Sigma +  \frac{\Theta}{3} \right)\Omega + \frac{1}{2}\varepsilon_{ab}\delta^a a^b,
\end{equation}

\begin{equation}
    \widehat{\zeta}_{\{ab\}} = \delta_{\{a}a_{b\}}+\left( \Sigma +  \frac{\Theta}{3} \right)\Sigma_{ab}-\mathcal{E}_{ab},
\end{equation}

\begin{equation}
  \begin{aligned}
    \widehat{V}_{\overline{a}}-\frac{2}{3}\widehat{W}_{\overline{a}}  & = -\delta_a\delta_b\Sigma^b-\varepsilon_{bc}\delta_a\delta^b\Omega^c-\frac{3\Sigma}{2}\delta_a\phi \\
    & \quad +2\left( \dot{\Sigma}-\frac{2}{3}\dot{\Theta} \right)\varepsilon_{ab}\Omega^b,
    \end{aligned}
\end{equation}

\begin{equation}
    \widehat{\Sigma}_{\overline{a}}-\varepsilon_{ab}\widehat{\Omega}^b = \frac{1}{2}V_a +\frac{2}{3}W_a-\varepsilon_{ab}\delta^b\Omega - \delta^b\Sigma_{ab} - \frac{3\Sigma}{2} a_a,
\end{equation}

\begin{equation}\label{OmegaVS}
    \widehat{\Omega} = -\delta_a\Omega^a,
\end{equation}

\begin{equation}
    \widehat{\Sigma}_{\{ab\}} = \delta_{\{a}\Sigma_{b\}}-\varepsilon_{c\{a}\delta^c\Omega_{b\}} - \varepsilon_{c\{a}\tensor{\mathcal{H}}{_{b\}}^c} + \frac{3\Sigma}{2}\zeta_{ab},
\end{equation}

\begin{equation}
    \widehat{X}_{\overline{a}}-\frac{1}{3}\widehat{\mu}_{\overline{a}} = -\delta_a\delta_b\mathcal{E}^b-\frac{3\mathcal{E}}{2}\delta_a\phi+2\left( \dot{\mathcal{E}}-\frac{2}{3}\dot{\mu} \right)\varepsilon_{ab}\Omega^b,
\end{equation}

\begin{equation}
    \widehat{\mathcal{E}}_{\overline{a}} = \frac{1}{2}X_a+\frac{1}{3}\mu_a-\delta^b\mathcal{E}_{ab}-\frac{3\mathcal{E}}{2}a_a-\frac{3\Sigma}{2}\varepsilon_{ab}\mathcal{H}^b,
\end{equation}

\begin{equation}
    \widehat{\mathcal{H}} = -\delta_a\mathcal{H}^a -(\mu+p+3\mathcal{E})\Omega,
\end{equation}

\begin{equation}
  \begin{aligned}
    \widehat{\mathcal{H}}_{\overline{a}} &  = \frac{1}{2}\delta_a\mathcal{H}-\delta^b\mathcal{H}_{ab}-\left( \mu+p -\frac{3\mathcal{E}}{2}\right) \Omega_a \\
    & \quad -\frac{3\mathcal{E}}{2}\varepsilon_{ab}\Sigma^b+\frac{3\Sigma}{2} \varepsilon_{ab}\mathcal{E}^b,
    \end{aligned}
\end{equation}

\begin{equation}
    \widehat{p} = -(\mu+p)\mathcal{A} \; .
\end{equation}

\noindent Lastly, the constraints are

\begin{equation}
    \delta_a\Omega^a+\varepsilon_{ab}\delta^a\Sigma^b = \mathcal{H}-3\Sigma\xi,
\end{equation}

\begin{equation}
    \frac{1}{2}\delta_a\phi - \varepsilon_{ab}\delta^b\xi-\delta^b\zeta_{ab} = \left( \frac{\Sigma}{2} - \frac{\Theta}{3} \right) \left( \varepsilon_{ab}\Omega^b-\Sigma_a \right) -\mathcal{E}_a,
\end{equation}

\begin{equation}
    V_a-\frac{2}{3}W_a+2\varepsilon_{ab}\delta^b\Omega+2\delta^b\Sigma_{ab} = -2\varepsilon_{ab}\mathcal{H}^b,
\end{equation}

\begin{equation}
    p_a = -(\mu+p)\mathcal{A}_a \; .
\end{equation}

\subsection{Harmonic Decomposition}
\label{subsectionharmonic}

To transform the linearized partial differential equations from the previous section into
ordinary differential equations in time we perform a harmonic decomposition of the
covariant quantities. The procedure is here shortly described and for more details
the reader is referred to \cite{Schperturb,Schperturb2,BFK,LRSIItensor}.  Different types of harmonics used in
relativity can, for example, be found in \cite%
{Challinor2,Gebbie1,Thorne,Harrison}.

Scalars are written as 
\begin{equation}
\Psi =\displaystyle\sum\limits_{k_{\parallel },k_{\perp }}\Psi
_{k_{\parallel }k_{\perp }}^{S}\ P^{k_{\parallel }}\ Q^{k_{\perp }}\ ,
\end{equation}%
where the harmonic coefficients $\Psi _{k_{\parallel }k_{\perp }}^{S}$ are functions of time.
Here the eigenfunctions $P^{k_{\parallel }}$ satisfy
\begin{equation}
\widehat{\Delta }P^{k_{\parallel }}\equiv n^{a}\nabla _{a}n^{b}\nabla _{b}=-\frac{k_{\parallel }^{2}}{a_{1}^{2}}%
P^{k_{\parallel }}\ ,\ \delta _{a}P^{k_{\parallel }}=\dot{P}^{k_{\parallel
}}=0\ ,
\end{equation}
where the constants $k_{\parallel }$ are dimensionless comoving wave numbers in the $n^a$ direction
and the physical wavenumbers are given by $k_{\parallel }/a_{1}$ in terms of the scale factor $a_1=a_1(t)$. 
In terms of the metric (\ref{metric}) the $P^{k_{\parallel }}$ can be represented by $e^{ik_{\parallel }z}$.
The eigenfunctions $Q^{k_\perp}$ satisfy
\cite{1+1+2}: 
\begin{equation}  \label{Beltrami}
\delta^2 Q^{k_\perp}\equiv \delta _{a}\delta ^{a} Q^{k_\perp}=-\frac{k_\perp^2}{a_2^2} Q^{k_\perp}\, ,\ \widehat{Q}%
^{k_{\perp }}=\dot{Q}^{k_{\perp }}=0\ ,
\end{equation}
where $a_{2}$ is the scale factor of
the 2-sheets, and $k_\perp$ are the dimensionless comoving wavenumbers along
the 2-sheets.
When ${\mathcal{R}}>0$ the 2-sheets are spheres and the eigenfunctions can be
represented by the spherical harmonics $Y_{l}^{m}$
with $k_{\perp }^{2}=l(l+1)$, $l=0,1,2,...$. Due to the background symmetry
the $m$-values will not appear explicitly in the equations.
When ${\mathcal{R}}\leq 0$, and the 2-sheets are open, the $k_\perp$ take continuous values.
For continuous values the sums go over into integrals.

Vectors $\Psi _{a}$ are expanded in terms of the even and odd vector harmonics  \cite{Schperturb,Schperturb2,LRSIItensor}
\begin{equation}\label{vectorharmonics}
Q_{a}^{k_{\perp }}=a_{2}\delta _{a}Q^{k_{\perp }}\ ,\ \ \overline{Q}%
_{a}^{k_{\perp }}=a_{2}\varepsilon _{ab}\delta ^{b}Q^{k_{\perp }}\ ,
\end{equation}%
as
\begin{equation}
\Psi _{a}=\displaystyle\sum\limits_{k_{\parallel },k_{\perp
}}P^{k_{\parallel }}\ \left( \Psi _{k_{\parallel }k_{\perp
}}^{V}Q_{a}^{k_{\perp }}+\overline{\Psi }_{k_{\parallel }k_{\perp }}^{V} 
\overline{Q}_{a}^{k_{\perp }}\right) \ .  \label{harmexpV0}
\end{equation}%
Note the factor $a_2$ appearing in the definitions of the vector harmonics in (\ref{vectorharmonics}),
giving the harmonic coefficients $\Psi _{k_{\parallel }k_{\perp}}^V$ and $\overline{\Psi }_{k_{\parallel }k_{\perp }}^{V}$
the right physical dimensions.

Similarly, a tensor $\Psi _{ab}$ can be expanded in terms of 
the even and odd tensor harmonics%
\begin{equation}
Q_{ab}^{k_{\perp }}=a_{2}^{2}\delta _{\{a}\delta _{b\}}Q^{k_{\perp }}\ ,\ 
\overline{Q}_{ab}^{k_{\perp }}=a_{2}^{2}\varepsilon _{c\{a}\delta ^{c}\delta
_{b\}}Q^{k_{\perp }}\ ,
\end{equation}%
as
\begin{equation}
\Psi _{ab}=\displaystyle\sum\limits_{k_{\parallel },k_{\perp
}}P^{k_{\parallel }}\ \left( \Psi _{k_{\parallel }k_{\perp
}}^{T}Q_{ab}^{k_{\perp }}+\overline{\Psi }_{k_{\parallel }k_{\perp }}^{T} 
\overline{Q}_{ab}^{k_{\perp }}\right) \ .
\end{equation}

Some properties of the vector and tensor harmonics are found
in appendix \ref{Qharmonics}. 

\subsection{Relations between harmonic coefficents}
\label{subsectionrelharmonics}

As mentioned above, the hat derivatives, $\hat G$, of the elements $G=\{\mu, p, {\mathcal{E}}, \Theta, \Sigma\}$
may be expressed in terms of $Z_a\equiv\delta_a G$. Expand $\hat G$ and $Z_a$ as
\begin{equation}\label{hatG}
\widehat{G} = \sum_{k_\parallel, k_\perp} \tilde{G}^S_{k_\parallel k_\perp} P^{k_\parallel}Q^{k_\perp} 
\end{equation}
and
\begin{equation}
Z_{a}=\displaystyle\sum\limits_{k_{\parallel },k_{\perp
}}P^{k_{\parallel }}\ \left( Z_{k_{\parallel }k_{\perp
}}^{V}Q_{a}^{k_{\perp }}+\overline{Z }_{k_{\parallel }k_{\perp }}^{V} 
\overline{Q}_{a}^{k_{\perp }}\right) \   \label{harmexpV0}
\end{equation}
respectively.
From the even part of the commutation relation (\ref{commutatorA3}) one then obtains
\begin{equation}\label{hatGcoefficient}
	\tilde{G}^S_{k_\parallel k_\perp}=2a_2\dot{G}\overline{\Omega}^V_{k_\parallel k_\perp}+\frac{ia_2k_\parallel }{a_1}Z^V_{k_\parallel k_\perp}  \ .
\end{equation}
Hence all $\hat G$ are given by (\ref{hatG}) with $\tilde{G}^S_{k_\parallel k_\perp}$ from (\ref{hatGcoefficient}).

Similarly, with the commutation relation (\ref{commutatorA4}) and the properties of the harmonics in appendix \ref{Qharmonics},
one can show that the odd coefficients of $Z_a$ satisfy
\begin{equation}
\overline{Z}^V_{k_\parallel k_\perp}=\frac{2a_2}{k_\perp ^2}\dot G\Omega^S_{k_\parallel k_\perp} \ .
\end{equation}

The degrees of freedom of the vorticity can also be decreased by substitution of the harmonic expansions of $\Omega$ and
$\Omega^a$ in equation (\ref{OmegaVS}), which gives
\begin{equation}
\Omega^V_{k_\parallel k_\perp} = \frac{ia_2k_\parallel}{a_1k_\perp^2}\Omega^S_{k_\parallel k_\perp} \ .
\end{equation}

In the following sections and the appendicies we will drop the indicies $k_\parallel  k_\perp$ on harmonic coefficient, since their
meaning will be obvious due to the superscripts $S$, $V$ and $T$.

\section{Evolution Equations}
\label{EvolutionEquations}

By decomposing the system of linearized equations given in section \ref{sectionperturbations} into harmonics we get a new system of scalar evolution equations and constraints, which can be found in appendix~\ref{LinearizedHarmonics}. The new system contains a total of 30 evolution equations, 28 constraints and 37 scalar variables, and it decouples into an even and odd sector. 

Before solving the new system of equations we can simplify it further. First, we note that we still have the freedom to partially fix the frame on the perturbed model, in choosing a direction for $n_a$. Based on the reasoning in \cite{GWKS}
 we choose $n_a$ such that $a_a=0$, which will simplify the system significantly. 
Also, we choose to look specifically on 
barotropic 
matter perturbations, which means that $p=p(\mu)$, and hence that $p^V=c_s^2\mu^V$, where $c_s^2=\dot{p}/\dot{\mu}$ is the square of the matter speed of sound.

With the choices above the even sector can now be solved in terms of the five free variables

\begin{equation}
    \label{evenfree}
	\left\{ \overline{\Omega}^V, \mu^V, \Sigma^T, \mathcal{E}^T, \overline{\mathcal{H}}^T \right\}
\end{equation}

\noindent and the odd sector in terms of the following three free variables, 

\begin{equation}
    \label{oddfree}
	\left\{ \Omega^S,  \overline{\mathcal{E}}^T, \mathcal{H}^T \right\}.
\end{equation}
Note that for vectors and tensors which are odd
by definition, like the magnetic part of the Weyl tensor, the r\^oles of quantities without and with an overbar are interchanged.
Hence, ${\mathcal{H}}^{T}$ belongs to the
odd sector, whereas $\overline{\mathcal{H}}^{T}$
belongs to the even sector.

It is of course possible to choose eight different variables than the ones above, though our choice is motivated by the results in \cite{GWKS, BFK}, where $\mathcal{E}$ and $\mathcal{H}$ represents gravitational waves. The remaining even variables, 

\begin{equation*}
	\left\{\phi^S, \mathcal{A}^S, \overline{\mathcal{H}}^V, \Sigma^V, \mathcal{E}^V, X^V, V^V, W^V, p^V, 
\mathcal{A}^V, \alpha^V,   \zeta^T \right\}
\end{equation*}

\noindent and odd variables

\begin{equation*}
  \begin{aligned}
	 \left\{ \xi^S, \right. &  \Omega^V, \mathcal{H}^S, \mathcal{H}^V, \overline{\Sigma}^V, \overline{\mathcal{E}}^V, \overline{X}^V, \overline{V}^V, \overline{W}^V, \\
 &  \overline{\mu}^V, \overline{p}^V,  
\overline{\mathcal{A}}^V, \overline{\alpha}^V, \overline{\Sigma}^T, \left. \overline{\zeta}^T \right\},
  \end{aligned}
\end{equation*}

\noindent can now all be expressed in terms of the previously introduced free variables. These expressions are found in appendix~\ref{HarmonicCoefficients}, and their consistency has been checked against the propagation equations found in appendix~\ref{LinearizedHarmonics}.

The equations of interest are now the evolution equations for the variables given in \eqref{evenfree} and \eqref{oddfree}. For the even sector we get the following equations,

\begin{equation}
	\label{eq467}
    \dot{\overline{\Omega}}^V = -\left(\left(\frac{2}{3}-c_s^2\right)\Theta+\frac{\Sigma}{2} \right)\overline{\Omega}^V,
\end{equation}

\begin{equation}\label{eqmu}
    \begin{aligned}
        \dot{\mu}^{V} & =-\left(\mu+p \right)\left(\frac{2i\kp}{a_1}-\frac{a_1 G}{ik_{\parallel }B} \right)\overline{\Omega}^V \\
        & \quad + \left( \frac{\Sigma }{2}\left( 1-\frac{3}{B} \left(\mu+p \right)\right) -\frac{4\Theta }{3}\right) \mu^{V}  \\ 
        & \quad +\frac{a_{2}}{2}\left( \mu +p\right) \left( \left( 1-C\right)\left( B\Sigma^{T}-3\Sigma \mathcal{E}^{T}\right)\vphantom{\frac{k_{\parallel}}{a_1}}  \right. \\ 
          & \quad \left. +\frac{ik_{\parallel }}{a_{1}}(2-J)\overline{\mathcal{H}}^{T}\right),
    \end{aligned}
\end{equation}

\begin{equation}\label{eqshear}
    \dot{\Sigma}^{T}=-\frac{c_{s}^{2}}{a_{2}\left(\mu +p\right) }\mu^{V}+\!\left( \Sigma -\frac{2\Theta }{3}\right) \Sigma^{T}-\mathcal{E}^{T},
\end{equation}

\begin{equation}\label{eqE}
  \begin{aligned}
    \dot{\mathcal{E}}^{T} & =-\frac{a_1 G}{ia_2k_{\parallel}B}\overline{\Omega}^V+\frac{3\Sigma }{2a_{2}B}\mu^{V}-\frac{\mu +p}{2}\Sigma^{T} \\
    & \quad -\frac{3}{2}\left( F+\Sigma C\right) \mathcal{E}^{T}-\frac{ik_{\parallel }}{2a_{1}}(2-J)\overline{\mathcal{H}}^{T},
    \end{aligned}
\end{equation}

\begin{equation}
    \label{eq67}
    \begin{aligned}
    \dot{\overline{\mathcal{H}}}^{T} & =\frac{2}{a_2 B}\left(\mu+p \right)\left(\Sigma+\frac{\Theta}{3}\right)\overline{\Omega}^V-\frac{ik_{\parallel }}{a_{1}a_{2}B}\mu^{V} \\ 
    & \quad -\frac{3}{2}\left( \frac{M}{B}+F\right) \overline{\mathcal{H}}^{T}-\frac{ik_{\parallel }}{a_{1}}\left( 1-C\right) \mathcal{E}^{T}.
    \end{aligned}
\end{equation}

\noindent Above, we have introduced the additional notations

\begin{equation}
	\tilde{k}^2 \equiv \frac{\okt}{a_2^2} + 2\frac{\pkt}{a_1^2},
\end{equation}

\begin{equation}
    B\equiv \tilde{k}^2+\frac{9\Sigma ^{2}}{2}+3\mathcal{E},
\end{equation}
\begin{equation}
    CB\equiv \Sigma\left(\Theta-\frac{3\Sigma}{2}\right)-\frac{k_\perp^2}{a_2^2},
\end{equation}
\begin{equation}
	G\equiv\left(\mu+p \right)\left(\mathcal{R}-\tilde{k}^2 \right),
\end{equation}
\begin{equation}
    JB\equiv \frac{k_\perp^2 a_1^2}{k_\parallel ^2 a_2^2}\left({\mathcal{R}}-\tilde{k}^2\right)+2\Sigma\left(\Theta-\frac{3\Sigma}{2}\right),
\end{equation}

\begin{equation}
    M\equiv 2{\mathcal{E}}\left( \Sigma +\frac{\Theta }{3}\right) +\Sigma \frac{{\mathcal{R}}a_{2}^{2}-k_{\perp }^{2}}{a_{2}^{2}}~,
\end{equation}

\begin{equation}
    F\equiv \Sigma + \frac{2\Theta}{3} \; .
\end{equation}

For the odd parity sector we get the following evolution equations,

\begin{equation}
    \label{eq75}
    \dot{\Omega}^S=\left(\Sigma+\left(c_s^2-\frac{2}{3}\right)\Theta \right)\Omega^S,
\end{equation}

\begin{equation}\label{eq76}
  \begin{aligned}
    \dot{\overline{\mathcal{E}}}^T &  = 
    P\Omega^S  
     -\frac{3}{2}\left(F+\Sigma D\right)\overline{\mathcal{E}}^{T}+\frac{ik_{\parallel }}{a_{1}}\left( 1-D\right)\mathcal{H}^{T},  
    \end{aligned}
\end{equation}

\begin{equation}
    \label{eq77}
    \begin{aligned}
        \dot{\mathcal{H}}^T & =  
        S\Omega^S 
        -\frac{a_{1}}{2ik_{\parallel }}\left( \frac{2k_{\parallel }^{2}}{a_{1}^{2}}-CB+9\Sigma E\right) \overline{\mathcal{E}}^{T} \\ 
        & \quad -\frac{3}{2}\left( 2E+F\right) \mathcal{H}^{T} 
    \end{aligned}
\end{equation}

\noindent where we have introduced

\begin{equation}
P\equiv  \frac{2}{k_\perp^2 B}\left(\mu+p\right)\left(\mu+p+3\mathcal{E}+B+\mathcal{R}-\frac{k_\perp^2}{a_2^2}\right),
\end{equation}

\begin{equation}
S \equiv \frac{2ia_1}{3k_\parallel k_\perp^2 B}\left(\mu+p \right)\left(3\Sigma \left( \frac{k_\perp^2}{a_2^2}-\frac{k_\parallel^2}{a_1^2}\right)+\Theta\tilde{k}^2 \right),
\end{equation}

\begin{equation}
    EB\equiv\frac{\Sigma}{2}\left(CB-\mathcal{E}\right)+\frac{\Theta\mathcal{E}}{3},
\end{equation}

\begin{equation}
    D\equiv C+\frac{\mu+p}{B} \; .
\end{equation}

Interestingly enough, the evolution of both the even and odd part of the vorticity completely decouples from the other variables, and hence we will treat it separately in the next section. The evolution equations for the remaining variables are consistent with previous research, see e.g. \cite{GWKS, BFK}, with the addition of the vorticity appearing as a source term for all variables except $\Sigma^T$. These equations are still quite cumbersome to solve analytically though, and so we will proceed to examine them in the high frequency limit in section~\ref{Optics}.

\section{Vorticity}\label{Vorticity}

As is well known both in the non-relativistic, cf. e.g. \cite{Raichoudhuri}, and relativistic cases, vorticity cannot be generated in a perfect fluid with barotropic equation of state $p=p(\mu)$, see e.g. \cite{Cargese, LuAnandaClarksonMaartens} for a proof in general relativity. Their argument is in short 
as follows:
in the 1+3 covariant split the evolution equation for the vorticity vector $\omega^a$ is given by 
\begin{equation}
h^{ab}\dot\omega_b=-\frac{2}{3}\Theta\omega^a+\sigma^a_{\;b}\omega^b+\frac{1}{2}\epsilon^{abc}D_b A_c
\end{equation}
\cite{Cargese}. Using the twice contracted Bianchi identities $D_a p+(\mu+p)A_a=0$ and $\dot \mu+\Theta(\mu+ p)$ 
and the commutator relation
\begin{equation}
D_{[a}D_{b]}\Psi=\epsilon_{abc}\omega^c\dot\Psi \, ,
\end{equation}
where $\Psi$ is an arbitrary scalar, it follows that
the last term can be rewritten as $c_s^2\Theta\omega^a$ when $p=p(\mu)$, so that
\begin{equation}\label{eqbaryrot}
h^{ab}\dot\omega_b=-\left(\frac{2}{3}-c_s^2\right)\Theta\omega^a+\sigma^a_{\;b}\omega^b \ .
\end{equation}
Hence there are no source terms, and since this equation is exact, vorticity cannot be generated to any order in a barotropic perfect fluid. 
For generation mechanisms due to entropy flow  
see e.g. \cite{Christopherson1,ChristophersonMalik} or from N-body simulations see \cite{CLAD}.

However, as seen from section \ref{EvolutionEquations}, an already existing vorticity will act as source terms in the evolution equations for the
density and electric and magnetic parts of the Weyl tensor, and may hence influence the growth of structures and give imprints on sonic and
gravitational waves.
For high frequency gravitational waves the effects of vorticity will be negligible though, as will be seen in section \ref{Optics}.

\subsection{Evolution of vorticity}
On projecting (\ref{eqbaryrot}) along $n^a$ and onto the 2-sheets with $N^a_{\; b}$ , using that $\omega^a=\Omega n^a+\Omega^a$, we obtain
\begin{equation}
 \dot{\Omega}=\left( \Sigma-\left(\frac{2}{3}-c_s^2\right)\Theta \right)\Omega+\left(\alpha_a+\Sigma_a\right)\Omega^a
\end{equation}
and 
\begin{equation}
 \dot{{\Omega}}^{\bar a} = -\left(\frac{\Sigma}{2}+\left(\frac{2}{3}-c_s^2\right)\Theta \right){\Omega}^a
+\Omega\left(\Sigma^a-\alpha^a\right) +\Sigma^a_{\; b}\Omega^b
\end{equation}
respectively. To first order we then have
\begin{equation}
 \dot{\Omega}=\left( \Sigma-\left(\frac{2}{3}-c_s^2\right)\Theta \right)\Omega
\end{equation}
and 
\begin{equation}\label{eqvecrot}
 \dot{{\Omega}}^{\bar a} = -\left(\frac{\Sigma}{2}+\left(\frac{2}{3}-c_s^2\right)\Theta \right){\Omega}^a \; .
\footnote{A scalar equation for $\Omega^a \Omega_a$ can be obtained by multiplying (\ref{eqvecrot}) with $\Omega_a$ giving
$ \left(\Omega^a{\Omega}_a\right){\dot{}} = -\left({\Sigma}+\left(\frac{4}{3}-2c_s^2\right)\Theta \right){\Omega}^a{\Omega}_a $
(for a scalar $\dot \Psi$ reduces to $\partial  \Psi/\partial t$ in the given background coordinates (\ref{metric})).}
\end{equation}
These are in agreement with the scalar equations (\ref{eq75}) and (\ref{eq467}) for the harmonic coefficents $\Omega^S$ and $\overline{\Omega}^V$ 
\begin{align}
\dot{\Omega}^S=\left( \Sigma-\left(\frac{2}{3}-c_s^2\right)\Theta \right)\Omega^S, \;  \\
    \dot{\overline{\Omega}}^V = -\left(\frac{\Sigma}{2}+\left(\frac{2}{3}-c_s^2\right)\Theta \right)\overline{\Omega}^V \, 
\end{align}
respectively.

By assuming a linear equation of state, $p=(\gamma-1)\mu$, and substituting equations (\ref{Thetabackgroundexp}) and (\ref{Sigmabackgroundexp}) for $\Theta$ and $\Sigma$ 
in terms of the scale factors $a_1$ and $a_2$, one can now easily solve
for either $\Omega$ and $(\Omega_a\Omega^a)^{1/2}$ or the corrresponding harmonic coefficients $\Omega^S$ and $\overline{\Omega}^V$
\footnote{Due to the definition of the vector harmonics, see (\ref{vectorharmonics}), the harmonic coefficients of the vorticity have the right dimension and time dependence.}.
The harmonic coefficients are given by
\begin{equation}
	\label{eq487}
	\Omega^S = C_\Omega a_1^{\gamma-1}a_2^{2(\gamma-2)}
\end{equation}

\noindent and

\begin{equation}
	\label{eq488}
	\overline{\Omega}^V = C_{\overline{\Omega}}a_1^{\gamma-2}a_2^{2\gamma-3}
\end{equation}
where $C_\Omega=C_{\Omega k_\parallel k_\perp}$ and $ C_{\overline{\Omega}}= C_{\overline{\Omega} k_\parallel k_\perp}$ are functions of
the harmonic numbers $k_\parallel$ and $k_\perp$, given by the initial conditions. 
For the dust case  ($\gamma=1$) and radiation  ($\gamma=4/3$) we then get
\begin{equation}
\Omega^S = C_\Omega a_2^{-2} \, , \quad \overline{\Omega}^V = C_{\overline{\Omega}}a_1^{-1}a_2^{-1}
\end{equation}
and
\begin{equation}
\Omega^S = C_\Omega a_1^{1/3} a_2^{-4/3} \, , \quad \overline{\Omega}^V = C_{\overline{\Omega}}a_1^{-2/3}a_2^{-1/3}
\end{equation}
respectively.
To compare with earlier results for the flat Friedmann universes \cite{Hawking} let us first consider the isotropic case with $a_1=a_2\equiv a$. Then
\begin{equation}
\Omega^S = C_\Omega a^{3\gamma-5}\quad \hbox{and} \quad \overline{\Omega}^V = C_{\overline{\Omega}}a^{3\gamma-5}
\end{equation}
so that for dust and radiation the vorticity goes as $a^{-2}$ and $a^{-1}$, in agreement with the results in \cite{Hawking}.

As a check of the result for the vorticity, we here consider the angular momentum to first order. Consider a small comoving volume element
with uniform angular velocity, which is proportional to the vorticity. Since it is a first order quantity, the other quantities 
can be treated to zeroth order. Note also that to this order the Lorentz factor is one. 
Due to the background isotropy around $n^a$, the component of the angular momentum along $n^a$ should be conserved in the general case, 
as all its components should in the fully isotropic case.

From equations (\ref{continuity}) and (\ref{Thetabackgroundexp}) with $p=(\gamma -1)\mu$ the equation
\begin{equation}
\mu = C_{\mu} a_1^{-\gamma}a_2^{-2\gamma}
\end{equation}
is obtained for some $C_{\mu}$. Volume increases as $V\sim a_1 a_2^2$ and radius perpendicular to the $n^a$ axis as $R \sim a_2$.
Hence the time dependence of the component of the angular moment parallel to $n^a$, $L^\parallel$, of a comoving element with mass $\tilde M$ goes as
\begin{equation}
  \begin{aligned}
L^\parallel & \sim \tilde M R^2 \Omega^\parallel \sim \mu V R^2 \Omega^S \\ 
    & \sim  a_1^{-\gamma}a_2^{-2\gamma}  a_1 a_2^2 a_2 ^2 a_1^{\gamma-1}a_2^{2(\gamma-2)}\sim 1 \, ,
  \end{aligned}
\end{equation}
i.e., it is independent of time. For rotation around an axis in the $xy$-plane perpendicular to $n^a$, we may choose this axis to be, say,  the $x$-axis due to the background isotropy. 
Then the radius perpendicular to the axis of rotation will go as ${\bf R}\sim a_1 {\hat{\bf n}}+a_2 {\hat{\bf  y}}$ (modulo non time-dependent coefficents in front of the
two different directions), so that the time dependence of $R^2 \sim a_1^2+a_2^2$.
The time dependence of the component of the angular moment perpendicular to $n^a$, $L^{\perp}$, then goes as
\begin{equation}
  \begin{aligned}
L^{\perp} & \sim \tilde M R^2 \Omega^{\perp} \sim \mu V R^2 \overline{\Omega}^V \\
    &\sim a_1^{-\gamma}a_2^{-2\gamma}  a_1 a_2^2\left(a_1^2+a_2^2\right)
a_1^{\gamma-2}a_2^{2\gamma-3} \\ 
 & \sim \frac{a_1}{a_2}+\frac{a_2}{a_1}
  \end{aligned}
\end{equation}
and hence $L^{\perp}$ is conserved only if $a_2 \sim\ a_1$, i.e. if the background is isotropic.

\subsection{Vorticity as source terms}\label{vorticityb}

From equations (\ref{eq75}), (\ref{eq76}) and (\ref{eq77}) for the odd parity sector we can derive the following inhomogeneous wave equations with damping for $\overline{\mathcal{E}}$ and ${\mathcal{H}}$
\begin{equation}\label{Eodd}
	\ddot{\overline{\mathcal{E}}}^T+q_{\overline{\mathcal{E}}_1}\dot{\overline{\mathcal{E}}}^T+q_{\overline{\mathcal{E}}_0}\overline{\mathcal{E}}^T=
s_{\overline{\mathcal{E}}_\Omega} \Omega^S,
\end{equation}
and
\begin{equation}\label{Hodd}
	\ddot{\mathcal{H}}^T+q_{\mathcal{H}_1}\dot{\mathcal{H}}^T+q_{\mathcal{H}_0}\mathcal{H}^T=s_{\mathcal{H}_\Omega}\Omega^S,
\end{equation}
where
\begin{align}\nonumber
	q_{\overline{\mathcal{E}}_0} & = \frac{3}{2}\left(F+\Sigma D\right)\frac{d}{dt}\ln\left(a_1\frac{F+\Sigma D}{1-D}\right)\\\nonumber
&+\frac{1}{2}\left(1-D\right)\left(\frac{2k_\parallel^2}{a_1^2}-BC+9\Sigma E\right)\\%\nonumber
&+\frac{9}{4}\left(F+\Sigma D\right)\left(F+2E\right),
 \\%\nonumber
q_{\overline{\mathcal{E}}_1} & = 3F+\frac{3}{2}\Sigma D+3E-\frac{d}{dt}\ln\frac{1-D}{a_1}, \\\nonumber
s_{\overline{\mathcal{E}}_\Omega} & =\frac{ik_\parallel}{a_1}\left(1-D\right)S+\Sigma P\\%\nonumber
	&+P\left[\frac{d}{dt}\ln\left(\frac{a_1 P}{1-D}\right)+\left(c_s^2-\frac{2}{3}\right)\theta+\frac{3}{2}\left(F+2E\right)\right],\\\nonumber
    q_{\mathcal{H}_0} & = \frac{3}{2}\left(2E+F\right)\frac{d}{dt}\ln\left(\frac{a_1(2E+F)}{\frac{2k_\parallel^2}{a_1^2}-BC+9\Sigma E}\right)\\\nonumber
&+\frac{1}{2}\left(1-D\right)\left(\frac{2k_\parallel^2}{a_1^2}-BC+9\Sigma E\right)\\%\nonumber
&+\frac{9}{4}\left(F+\Sigma D\right)\left(F+2E\right),\\\nonumber
 q_{\mathcal{H}_1} & = 3F+3E+\frac{3}{2}\Sigma D\\%\nonumber
 &-\frac{d}{dt}\ln\left(a_1\left(\frac{2k_\parallel^2}{a_1^2}-BC+9\Sigma E\right)\right),\\\nonumber
s_{\mathcal{H}_\Omega}& = -\frac{a_1}{2ik_\parallel}\left(\frac{2k_\parallel^2}{a_1^2}-BC+9\Sigma E\right)P+\Sigma S\\
&+S\left[\frac{d}{dt}\ln\left(\frac{a_1S}{\frac{2k_\parallel^2}{a_1^2}-BC+9\Sigma E}\right)+\left(c_s^2-\frac{2}{3}\right)\theta\right].
\end{align}
Hence vorticity, which as seen is independent of the other quantities, acts as a source for the $\overline{\mathcal{E}}$ and ${\mathcal{H}}$ perturbations. 
However, as will be seen in section \ref{Optics}, for large
wave numbers, $k_{\parallel},k_{\perp}$, the source terms become negligible.

Similarly we can from the even parity equations (\ref{eq467}), (\ref{eqmu}), (\ref{eqshear}), (\ref{eqE}) and (\ref{eq67}) derive wavelike equations sourced by
the density gradient $\mu^V$, its time derivative $\dot\mu^V$ and the vorticity component $\overline{\Omega}^V$ for the
shear $\Sigma ^T$ and the components ${\mathcal{E}^T}$ and $\overline{\mathcal{H}}^T$ of the Weyl tensor. Note 
however that to obtain pure equations in  $\Sigma ^T$, ${\mathcal{E}^T}$  or $\overline{\mathcal{H}}^T$, sourced only by the
vorticity, one needs to go to higher order differential equations than second order.

For the shear one gets
\begin{eqnarray}\nonumber
\ddot{{\Sigma}}^T+\left(2\Sigma+\frac{5}{3}\Theta\right)\dot{{\Sigma}}^T+q_{{\Sigma}_0}{\Sigma}^T=\\\label{Sheareven}
\frac{2ik_\parallel}{a_1 a_2}\overline{\Omega}^V+\frac{1-c_s^2}{a_2(\mu+p)}\dot\mu^V+s_{\Sigma\mu}\mu^V,
\end{eqnarray}

\noindent where

\begin{eqnarray}\nonumber
q_{{\Sigma}_0}&= &\frac{(1-C)B-3(\mu+p)}{2}-\\
&&\frac{3}{2}(3\Sigma+2\Theta)\left(\Sigma-\frac{2\Theta}{3}\right),\\\nonumber
s_{\Sigma \mu}&=&\left[\frac{4\Theta}{3}-\frac{\Sigma}{2}-c_s^2(\Theta+3\Sigma)-\frac{d}{dt} (c_s^2)\right.\\\label{eqsSigmamu}
&&\left.+c_s^2\frac{d}{dt}\ln\left(a_2(\mu+p)\right)\right]\frac{1}{a_2(\mu+p)}\, ,
\end{eqnarray}

\noindent for the magnetic part of the Weyl tensor

\begin{equation}\label{Heven}
	\ddot{\overline{\mathcal{H}}}^T+q_{\overline{\mathcal{H}}_1}\dot{\overline{\mathcal{H}}}^T+q_{\overline{\mathcal{H}}_0}\overline{\mathcal{H}}^T=
s_{\overline{\mathcal{H}}_{\Omega}}\overline{\Omega}^V+s_{\overline{\mathcal{H}}_{\mu}}\mu^V,
\end{equation}

\noindent where
\begin{align}\nonumber
    q_{\overline{\mathcal{H}}_0} & =\frac{3}{2}\left(\frac{M}{B}+F\right)\left[\frac{3}{2}\left(\frac{\mu+p}{B}\Sigma+F+\Sigma C\right)-\right.\\\nonumber
& \left.   \frac{d}{dt}\ln\left(\frac{1-C}{a_1\left(\frac{M}{B}+F\right)}\right)\right] \\%\nonumber
 & +\frac{k_\parallel^2}{2a_1^2}(2-J)\left(1-C-\frac{\mu+p}{B}\right),\\\nonumber
 q_{\overline{\mathcal{H}}_1} & = \frac{3}{2}\left(\frac{M}{B}+2F+\Sigma C+\frac{(\mu+p)}{B}\Sigma\right) \\%\nonumber
 &-\frac{d}{dt}\ln\left(\frac{1-C}{a_1}\right),\\%\nonumber
s_{\overline{\mathcal{H}}_{\mu}} & =\frac{ik_\parallel}{a_1a_2B}\left(\frac{\Theta}{3}-\frac{7\Sigma}{2}+\frac{d}{dt}\ln\left(a_2B(1-C)\right)\right),\\\nonumber
s_{\overline{\mathcal{H}}_{\Omega}} & =\frac{1}{a_2B}\left(G(1-C)-(\mu+p)\left(\frac{2k_\parallel^2}{a_1^2}-\frac{G}{B}\right)\right)-\\\nonumber
& \left(\frac{G}{B}+\mu+p\right)\frac{1}{a_2 \Sigma}\left[\Sigma C +\frac{2}{3}\left((3c_s^2+1)\frac{\Theta}{3}+\Sigma\right)+\right.\\
& \left. \frac{(\mu+p)\Sigma}{B}+\frac{2}{3}\frac{d}{dt}\ln\left(\frac{\left(\frac{G}{B}+\mu+p\right)a_1}{a_2\Sigma(1-C)}\right)\right]\, ,
\end{align}
and for the electric part of the Weyl tensor
\begin{equation}\label{Eeven}
	\ddot{{\mathcal{E}}}^T+q_{{\mathcal{E}}_1}\dot{{\mathcal{E}}}^T+q_{{\mathcal{E}}_0}{\mathcal{E}}^T=
s_{{\mathcal{E}}_\Omega} \overline{\Omega}^V+s_{{\mathcal{E}}_{\dot\mu}}\dot\mu^V+s_{{\mathcal{E}}_\mu} \mu^V,
\end{equation}
where
\begin{align}\nonumber
	q_{\mathcal{E}_0} & = \frac{3}{2}\frac{d}{dt}\left( \Sigma +\frac{2\Theta }{3}%
+\Sigma C\right) +\frac{\left( 1-C\right) k_\parallel^2\left(2-J\right)}{2 a_1^2}-  \notag \\
&W_{1}\!\left( \Theta +\frac{3}{2}\Sigma \left( 1+C\right) \right) \left[ 
\frac{d}{dt}\ln \left(\frac{2-J}{a_1}\right) -\frac{3M}{2B}-\frac{3F}{2}\right]  \notag \\
&+W_{2}\left( \!\Theta +3\Sigma \right) \left[ \Sigma -\frac{2\Theta }{3}+%
\frac{d}{dt}\ln \left( \mu +p\right) \right] -\frac{\mu +p}{2}  \notag \\
&-W_{2}\frac{3\Sigma \left( 1-C\right) }{2}\left[ \frac{d}{dt}\ln \left(
\frac{2-J}{a_1}\right) - \frac{3M}{2B}-\frac{3F}{2}\right] \!~, \\% \notag \\
q_{\mathcal{E}_1} & = 
-W_{2}\left[ \frac{5\Theta }{3}+\frac{\Sigma }{2}\left( 1+3C\right) -\frac{%
d}{dt}\ln \left( \mu +p\right) \right]+ ~, \notag  \\
& W_{1}\left[  \frac{3M}{2B}+\frac{3F}{2}\!\!+\!\Theta +\frac{3}{2}\Sigma
\left( 1+C\right) -\frac{d}{dt}\ln \left(\frac{2-J}{a_1}\right) \right] \, , \\%\notag \\
s_{{\mathcal{E}}_\mu} & = \frac{d}{dt}\!\left( \frac{3\Sigma }{2a_{2}B}\right) -%
\frac{k_\parallel^2\left(2-J\right)}{2a_{2} a_1^2 B}+\frac{c_{s}^{2}}{2a_{2}}  \notag \\
&+\frac{W_{3}-W_{4}}{a_{2}\left[ \left( 1-C\right) \!B\!-\left( \mu
+p\right) \right] }\!~, \\% \notag \\  
s_{{\mathcal{E}}_{\dot\mu}} & =\frac{3\Sigma }{2Ba_2}\!\!+\frac{1}{a_2\left[\left( 1-C\right)
\!B\!-\left( \mu +p\right)\right] }\times  \notag \\
&\left(\frac{d}{dt}\ln\left( \frac{%
2-J}{a_1(\mu +p)}\right) -\frac{\Theta }{3}-\frac{5\Sigma}{2} -\frac{3M}{2B}\right) ~, \\ % \notag \\
s_{{\mathcal{E}}_\Omega} & =\frac{a_1G}{ik_\parallel a_2 B}\left[\left(\frac{2}{3}-c_s^2\right)\Theta
\frac{\Sigma}{2}-\frac{d}{dt}\ln\left(\frac{a_1 G}{a_2 B}\right)\right] \notag \\
&+\frac{2ik_\parallel\left[\frac{d}{dt}\ln\left(\mu+p\right)+\Sigma-\frac{2\Theta}{3}\right]}
{a_1 a_2 \left(\mu+p\right)\left[\mu+p-\left(1-C\right)B\right]}+\notag \\
&\left(\frac{2ik_\parallel\left(\mu+p\right)}{a_1 a_2 \left[\mu+p-\left(1-C\right)B\right]}-\frac{a_1 G}{ia_2k_\parallel B}\right)\times \notag \\
&\left(\frac{3}{2}\left(\frac{M}{B}+F\right)-\frac{d}{dt}\ln\left(\frac{2-J}{a_1}\right)\right)
\end{align}
with the definitions
\begin{eqnarray}
W_{1} &\equiv&\frac{\left( 1-C\right) \!B}{\left( 1-C\right) \!B\!-\left( \mu
+p\right) }~, \\ %\notag \\
W_{2} &\equiv&\frac{\left( \mu +p\right) }{\left( 1-C\right) \!B\!-\left( \mu
+p\right) }~,\\ %  \notag \\
W_{3} &\equiv&\left[ \frac{d}{dt}\ln \left( \frac{2-J}{a_1}\right)  -\frac{3M}{2B}-\frac{3F}{2}\right] 
\notag \\
&&\times \left[ \frac{4\Theta }{3}-\!\frac{\Sigma \left( 4-3C\right) \!}{2}%
+\!\frac{3\Sigma \left( \mu +p\right) }{2B}\right] ~,\\ %  \notag \\
W_{4} &\equiv&\left[ \frac{d}{dt}\ln \left( \mu +p\right) +\Sigma -\frac{%
2\Theta }{3} \right] \left( \frac{4\Theta }{3}-\frac{\Sigma }{2}\right). \label{WQ}
\end{eqnarray}
Their behaviour for large wave numbers will be considered in section \ref{Optics}.

\section{Geometrical Optics Approximation}
\label{Optics}

We now turn our attention to  
the high frequency limit, also known as the geometrical optics approximation. Following the definitions given in \cite{Isaacson1,Isaacson2}, 
our current variables can in the limit 
be expressed as 

\begin{equation}
	\frac{\pkt}{a_1^2}, \frac{\okt}{a_2^2} \gg \Theta^2, \Sigma^2, \mathcal{E}, \mu, p\, ,
\end{equation}
where the physical wave numbers along the direction of anisotropy are given by $k_{\parallel}/a_1$ and along the 2-surfaces by $k_{\perp}/a_2$. Hence  
we define the total physical wave number $k$ as
\begin{equation}
	k^2 \equiv \frac{\okt}{a_2^2} + \frac{\pkt}{a_1^2}\, .
\end{equation}
\noindent 
Our results will now be compared to previous results, 
see \cite{GWKS} and \cite{BFK}, where the only differences 
should be related to the impact of $\Omega^S$ and $\overline{\Omega}^V$. As we saw in the previous section though, the evolution of the vorticity completely decouples from the other variables, independent of the magnitude of the frequency. However, we will see that the vorticity appears as a first order source term in the evolution equation for the shear.

We now rewrite the second order equations from section \ref{vorticityb}
in the form of the general wave equation for a time dependent scalar $X$ 
\begin{equation}
	\ddot{X}+2\zeta\omega\dot{X}+\omega^2X=Z
\end{equation}
\noindent where $\zeta$ represents the damping parameter, $\omega$ represents the undamped angular frequency and $Z$ is some time dependent factor which acts as a source term, producing forced oscillations. On assuming that the change of the scale factors are negligible over one period, we have that 
the real angular frequency is obtained from $\omega\sqrt{1-\zeta^2}$ with $|\zeta| < 1$, which means that the propagation speed of the wave is given by $\omega\sqrt{1-\zeta^2}/k$.

First we consider the even sector, given by the  equations \eqref{eqmu}, \eqref{eqshear},  \eqref{eqE} and  \eqref{eq67}.
By keeping only the highest order terms in $k$ they produce undamped wave equations, where the density waves $\mu^V$ and the shear waves $\Sigma^T$ progagate with
the speed of sound $c_s$, and the Weyl tensor components ${\mathcal{E}}^T$ and ${\overline{\mathcal{H}}}^T$ propagate as gravitational waves with
the speed of light.

 To next order in $k$ also damping and source terms are introduced. The equations \eqref{Eeven} and \eqref{Heven} for
${\mathcal{E}}^T$ and ${\overline{\mathcal{H}}}^T$ then reduce to
\begin{equation}
    \label{eq101}
	\ddot{\mathcal{E}}^T+q_{\mathcal{E}_1}\dot{\mathcal{E}}^T+k^2\mathcal{E}^T=\frac{1}{2a_2}\left(c_s^2-1\right)\mu^V,
\end{equation}
\begin{equation}
    \label{eq102}
	\ddot{\overline{\mathcal{H}}}^T+q_{\overline{\mathcal{H}}_1}\dot{\overline{\mathcal{H}}}^T+k^2\overline{\mathcal{H}}^T=0,
\end{equation}

\noindent where

\begin{align}
	q_{\mathcal{E}_1}  &=
                                             \frac{7\Theta}{3}+4\Sigma-6\Sigma\frac{k_\perp^2}{\tilde k^2 a_2^2}, \\
    q_{\overline{\mathcal{H}}_1}&  =    \frac{7\Theta}{3}+4\Sigma-3\Sigma\frac{k_\perp^2}{ k^2 a_2^2}       \, . 
\end{align}

Hence, $\overline{\mathcal{H}}^T$ and 
$\mathcal{E}^T$ represent damped gravitational waves which to 
zeroth and first order in $1/k$ 
propagates at the speed of light. 
This can be seen as follows: From equation (\ref{eq101}) for ${\mathcal{E}}^T$ 
we can read off the undamped angular frequency $\omega$ to be $k$. Hence the propagation velocity becomes $\sqrt{1-\zeta^2}$ and $\zeta$ will be given by $q_{\mathcal{E}_1}/2k$. Expansion of this gives that $\zeta$ is of order $1/k$, so that the propagation velocity is $1+{\mathcal{O}}(1/k^2)$. Similarly, 
${\overline{\mathcal{H}}}^T$ also propagate with the speed of light up to first order. 

Note that $\mu^V$, due to its definition as a gradient, in itself
is one order higher in $k$ than the other quantities. This can also be seen from equations \eqref{eqmu} and \eqref{eqshear}. The leading coefficient on the right hand side in \eqref{eqmu}
goes as $k^2$, whereas the leading coefficient in \eqref{eqshear} goes as $k^0$. On noting that each time derivative picks out a factor of order $k$, we
then for consistency need $\mu^V$ to be one order higher in $k$ than $\Sigma^T$.
Hence equation \eqref{eq101} for the electric part of the Weyl tensor ${\mathcal{E}}^T$ is found to be sourced by the density gradient $\mu^V$ to first order.
\footnote{This term is missing in the corresponding equations (5.29) and (79) in references \cite{GWKS} and \cite{BFK} respectively.}

The shear equation \eqref{Sheareven} becomes
\begin{eqnarray}
	\label{eqwavesigma}\nonumber
	&&\ddot{\Sigma}^T+\left(2\Sigma+\frac{5\Theta}{3}\right)\dot{\Sigma}^T+k^2\Sigma^T=\\
          &&\frac{2ik_\parallel}{a_1 a_2}\overline{\Omega}^V+\frac{1-c_s^2}{a_2(\mu+p)}\dot\mu^V+s_{\Sigma_{\mu}}\mu^V \, ,
\end{eqnarray}
where $s_{\Sigma_{\mu}}$ is given by \eqref{eqsSigmamu},
and the evolution equation \eqref{eqmu} for the density gradient becomes
\begin{equation}\label{eqmuhigh}
    \begin{aligned}
	\dot{\mu}^V &  =  a_2k^2\left(\mu+p \right)\Sigma^T  - \frac{a_2a_1\tilde{k}^2}{2i\kp}\left(\mu+p \right)\overline{\mathcal{H}}^T \\
      & \quad- \frac{2i\kp}{a_1}\left(\mu+p \right)\overline{\Omega}^V + \left(\frac{\Sigma}{2}-\frac{4\Theta}{3}\right)\mu^V\, 
    \end{aligned}
\end{equation}
on keeping terms to the two highest orders in $k$.  
Substitution of \eqref{eqmuhigh} in \eqref{eqwavesigma} gives
\begin{eqnarray}
	\label{eqwavesigma}\nonumber
	&&\ddot{\Sigma}^T+\left(2\Sigma+\frac{5\Theta}{3}\right)\dot{\Sigma}^T+c_s^2k^2\Sigma^T=\\
          &&\frac{2ik_\parallel c_s^2}{a_1 a_2}\overline{\Omega}^V+\left(c_s^2-1 \right)\frac{a_1\tilde{k}^2}{2i\pk}\overline{\mathcal{H}}^T+s_{\Sigma_{\mu}}\mu^V \, .
\end{eqnarray}
The density gradient $\mu^V$ on the right hand side can be eliminated by using \eqref{eqshear}, which to leading order is given by
\begin{equation}\label{eqmuk2}
\mu^V = -\frac{a_2\left(\mu+p\right)}{c_s^2}\dot\Sigma^T \, .
\end{equation}

Substitution of \eqref{eqmuk2} into \eqref{eqwavesigma} gives
\begin{eqnarray}
	\label{eq512}\nonumber
&&	\ddot{\Sigma}^T+\left(\left(\frac{4}{3}-c_s^2\right)\Theta-2\Sigma-\frac{d\left(c_s^2\right)}{dt}/c_s^2\right)\dot{\Sigma}^T+c_s^2k^2\Sigma^T\\
&&=\left(c_s^2-1 \right)\frac{a_1\tilde{k}^2}{2i\pk}\overline{\mathcal{H}}^T + c_s^2\frac{2i\pk}{a_1a_2}\overline{\Omega}^V
\end{eqnarray}
where $\overline{\mathcal{H}}^T$ and $\overline{\Omega}^V$ are determined from \eqref{eq102} and \eqref{eq467} respectively.
Hence the shear waves propagate as damped 
\footnote{Note that the damping factor differs from that for the corresponding equation (5.32) in \cite{GWKS}.}
sound waves with  $\overline{\Omega}^V$ and $\overline{\mathcal{H}}^T$ appearing as source terms
with coefficients of order $k$, i.e. they are corrections of the same order as the damping.
The last variable $\mu^V$ represent density waves which also propagate at the speed of sound, but
$\pi/2$ out of phase relative to the shear waves, 
as can be seen from \eqref{eqmuk2}.

In the same manner the evolution equations for the odd parity sector, equation \eqref{Eodd} and \eqref{Hodd}, can be rewritten as

\begin{equation}
	\ddot{\overline{\mathcal{E}}}^T+q_{\overline{\mathcal{E}}_1}\dot{\overline{\mathcal{E}}}^T+k^2\overline{\mathcal{E}}^T=0,
\end{equation}

\begin{equation}
	\ddot{\mathcal{H}}^T+q_{{\mathcal{H}}_1}\dot{\mathcal{H}}^T+k^2\mathcal{H}^T=0,
\end{equation}

\noindent where

\begin{align}
	q_{\overline{\mathcal{E}}_1}  &=   \frac{7\Theta}{3}+4\Sigma-3\Sigma\frac{k_\perp^2}{ k^2 a_2^2}, \\
    q_{\mathcal{H}_1} & =     \frac{7\Theta}{3}+4\Sigma-6\Sigma\frac{k_\perp^2}{\tilde k^2 a_2^2} .
\end{align}

\noindent Since there is no appearance of the vorticity in these equations the result is exactly the same as the one obtained in \cite{GWKS, BFK} where the vorticity vanishes on the perturbed model. Hence, both $\overline{\mathcal{E}}^T$ and $\mathcal{H}^T$ represents decoupled gravitational waves, which will to first order propagate at the speed of light.

\section{Conclusions}\label{conclusions}
\noindent In this paper we have presented a general treatment of perturbations of orthogonal
LRS class II cosmological backgrounds with non-vanishing vorticity. We have utilized the 1+1+2 covariant decomposition of spacetime and a harmonic decomposition of the resulting quantities, and the result is given as time evolution equations for the following eight harmonic coefficients, which decouple into an even sector

\begin{equation}
    \label{evenfree}
	\left\{ \overline{\Omega}^V, \mu^V, \Sigma^T, \mathcal{E}^T, \overline{\mathcal{H}}^T \right\}
\end{equation}

\noindent and and odd sector
\begin{equation}
    \label{oddfree}
	\left\{ \Omega^S,  \overline{\mathcal{E}}^T, \mathcal{H}^T \right\}.
\end{equation}

\noindent The remaining 27 harmonic coefficients can be desribed by the eight given above, which then completely describes the full model in a gauge-invariant way, utilizing the Stewart-Walker lemma.

We found that the evolution of the vorticity decouples from the other variables, as is expected, and given a linear equation of state it can be solved for analytically. The vorticity also appears as a source term for the other free variables. In the high frequency limit the free variables are found to represent graviational, density and shear waves, and the vorticity source terms disappears for all variables except the shear and density waves.

A natural 
extension to the work presented in this article would be to introduce vorticity creating mechanisms by considering, e.g., imperfect fluid perturbations with
viscosity and heat flow.
This would imply the introduction of 
more terms in 
the energy momentum tensor $T_{ab}$, one symmetric and trace-free 3-tensor $\pi^{ab}$ and
one spacelike vector $q^a$, 
together with coefficients of dynamic and bulk viscosities and of heat conductivity.
For the exact equations with a general energy momentum tensor in the 1+1+2 formalism see \cite{1+1+2},
and for the theory of relativistic imperfect fluids \cite{Eckart,Israel,Carter}.
Since a heat flow alone cannot generate vorticity to first order in perturbation theory, \cite{Christopherson1,ChristophersonMalik},
it would then be of interest to consider also 
second order 
perturbations. 
For works on second order perturbations in covariant perturbation theory see,
e.g. \cite{Osano2017}.
\vspace{6pt}

%\section*{Acknowledgements}

\appendix

\section{Commutation Relations}

\label{commutation} 

\noindent For a zeroth-order scalar field $\Psi$ on orthogonal and homogenous LRS class II backgrounds, with $\phi=0$, the following first order commutation relations hold:

\begin{equation}
    \widehat{\dot{\Psi}}-\dot{\widehat{\Psi }}=-\mathcal{A}\dot{\Psi}+\left(\Sigma +\frac{\Theta }{3}\right) \widehat{\Psi },
\end{equation}

\begin{equation}
    \delta _{a}\dot{\Psi}-N_{a}^{\,\,\,b}\left( \delta _{b}\Psi \right) ^{\cdot}=-\mathcal{A}_{a}\dot{\Psi}-\frac{1}{2}\left( \Sigma -\frac{2\Theta }{3}\right) \delta _{a}\Psi,
\end{equation}

\begin{equation}\label{commutatorA3}
    \delta _{a}\widehat{\Psi }-N_{a}^{\,\,\,b}\left( \widehat{\delta _{b}\Psi }\right) =-2\varepsilon _{ab}\Omega ^{b}\dot{\Psi} , 
\end{equation}

\begin{equation}\label{commutatorA4}
    \delta _{\lbrack a}\delta _{b]}\Psi =\varepsilon _{ab}\Omega \dot{\Psi}\ .
\end{equation}

For a first-order 2-vector $\Psi _{a}$ the following commutation relations, to first order, hold:

\begin{equation}
    \widehat{\dot{\Psi}}_{\bar{a}}-\dot{\widehat{\Psi }}_{\bar{a}}=\left( \Sigma+\frac{\Theta }{3}\right) \widehat{\Psi }_{\bar{a}}\ ,
\end{equation}

\begin{equation}
    \delta _{a}\dot{\Psi}_{b}-N_{a}^{\,\,\,c}N_{b}^{\,\,\,d}\left( \delta_{c}\Psi _{d}\right) ^{\cdot}=-\frac{1}{2}\left( \Sigma -\frac{2\Theta }{3}\right) \delta _{a}\Psi _{b}\ ,
\end{equation}

\begin{equation}
    \delta _{a}\widehat{\Psi }_{b}-N_{a}^{\,\,\,c}N_{b}^{\,\,\,d}\left( \widehat{\delta _{c}\Psi _{d}}\right)=0 ,
\end{equation}

\begin{equation}
    \delta _{\lbrack a}\delta _{b]}\Psi _{c}=\frac{1}{2}\mathcal{R}N_{c[a}\Psi_{b]}\ ,
\end{equation}

\noindent and for a first-order trace-free and symmetric 2-tensor $\Psi _{ab}$ it holds that:

\begin{equation}
    \widehat{\dot{\Psi}}_{\{ab\}}-\dot{\widehat{\Psi }}_{\{ab\}}=\left( \Sigma +\frac{\Theta }{3}\right) \widehat{\Psi }_{\bar{a}\bar{b}}\ ,
\end{equation}

\begin{equation}
    \delta _{a}\dot{\Psi}_{bc}-N_{a}^{\,\,\,d}N_{b}^{\,\,\,e}N_{c}^{\,\,\,f}\left( \delta _{d}\Psi _{ef}\right) ^{\cdot }=-\frac{1}{2}\left( \Sigma -\frac{2\Theta }{3}\right) \delta _{a}\Psi _{bc}\ ,
\end{equation}

\begin{equation}
    \delta _{a}\widehat{\Psi }_{bc}-N_{a}^{\,\,\,d}N_{b}^{\,\,\,e}N_{c}^{\,\,\,f}\left( \widehat{\delta _{d}\Psi _{ef}}\right) =0 ,
\end{equation}

\begin{equation}
    2\delta _{\lbrack a}\delta _{b]}\Psi _{cd}=\mathcal{R}\left( N_{c[a}\Psi_{b]d}+N_{d[a}\Psi _{b]c}\right).
\end{equation}

\section{Harmonics}\label{Qharmonics}

For both the vector and tensor spherical harmonics there exists orthogonality, algebraic and differential relations that are essential when decomposing the linearized 1+1+2 equations in terms of harmonic coefficients. For the vector spherical harmonics, the orthogonality relations are

\begin{equation}
	\tensor{N}{^a^b}Q_a^{k_\perp}\overline{Q}_b^{k_\perp}=0,
\end{equation}

\noindent the algebraic relations are

\begin{equation}
	Q_a^{k_\perp}=-\tensor{\varepsilon}{_a^b}\overline{Q}_b^{k_\perp}, \quad \overline{Q}_a^{k_\perp}=\tensor{\varepsilon}{_a^b}Q_b^{k_\perp}
\end{equation}

\noindent and the differential relations are

\begin{equation}
	\dot{Q}_a^{k_\perp}=\widehat{Q}_a^{k_\perp}=0, \quad \dot{\overline{Q}}_a^{k_\perp}=\widehat{\overline{Q}}_a^{k_\perp}=0,
\end{equation}

\begin{equation}
	\delta^2Q_a^{k_\perp}=\frac{\mathcal{R}a_2^2-2\okt}{2a_2^2} Q_a^{k_\perp}, \quad \delta^2\overline{Q}_a^{k_\perp}=\frac{\mathcal{R}a_2^2-2\okt}{2a_2^2} \overline{Q}_a^{k_\perp},
\end{equation}

\begin{equation}
	\delta^a Q_a^{k_\perp} = - \frac{\okt}{a_2}Q^{k_\perp}, \quad \delta^a\overline{Q}_a^{k_\perp}=0,
\end{equation}

\begin{equation}
	\varepsilon^{ab}\delta_a Q_b^{k_\perp}=0, \quad \varepsilon^{ab}\delta_a \overline{Q}_b^{k_\perp}=\frac{\okt}{a_2}Q^{k_\perp}.
\end{equation}

For the tensor spherical harmonics the orthogonality relations are

\begin{equation}
	N^{ab}N^{cd}Q_{ac}^{k_\perp}\overline{Q}_{bd}^{k_\perp}=0,
\end{equation}

\noindent the algebraic relations are

\begin{equation}
	Q_{ab}^{k_\perp}=\tensor{\varepsilon}{_{\{a}^c}\overline{Q}_{b\}c}^{k_\perp}, \quad \overline{Q}_{ab}^{k_\perp}=-\tensor{\varepsilon}{_{\{a}^c}Q_{b\}c}^{k_\perp}, 
\end{equation}

\noindent and the differential relations are

\begin{equation}
	\dot{Q}_{ab}^{k_\perp}=\widehat{Q}_{ab}^{k_\perp}=0, \quad \dot{\overline{Q}}_{ab}^{k_\perp}=\widehat{\overline{Q}}_{ab}^{k_\perp}=0,
\end{equation}

\begin{equation}
	\delta^2 Q_{ab}^{k_\perp}=\frac{2\mathcal{R}a_2^2-k_\perp^2}{a_2^2}Q_{ab}^{k_\perp}, \quad \delta^2\overline{Q}_{ab}^{k_\perp}=\frac{2\mathcal{R}a_2^2-k_\perp^2}{a_2^2}\overline{Q}_{ab}^{k_\perp},
\end{equation}

\begin{equation}
	\delta^b Q_{ab}^{k_\perp} = \frac{\mathcal{R}a_2^2-k_\perp^2}{2a_2}Q_a^{k_\perp}, \quad \delta^b \overline{Q}_{ab}^{k_\perp} = -\frac{\mathcal{R}a_2^2-k_\perp^2}{2a_2}\overline{Q}_a^{k_\perp},
\end{equation}

\begin{equation}
	\tensor{\varepsilon}{_a^c}\delta^bQ_{bc}^{k_\perp}=\frac{\mathcal{R}a_2^2-k_\perp^2}{2a_2}\overline{Q}_a^{k_\perp},  \tensor{\varepsilon}{_a^c}\delta^b\overline{Q}_{bc}^{k_\perp}=\frac{\mathcal{R}a_2^2-k_\perp^2}{2a_2}Q_a^{k_\perp},
\end{equation}

\begin{equation}
	\tensor{\varepsilon}{^b^c}\delta_bQ_{ac}^{k_\perp}=\frac{\mathcal{R}a_2^2-k_\perp^2}{2a_2}\overline{Q}_a^{k_\perp},  \tensor{\varepsilon}{^b^c}\delta_b\overline{Q}_{ac}^{k_\perp}=\frac{\mathcal{R}a_2^2-k_\perp^2}{2a_2}Q_a^{k_\perp}.
\end{equation}

\section{Harmonic Expansion of Linearized Equations}
\label{LinearizedHarmonics}

Here the harmonic expansion of the first order equations given in section \ref{sectionLE} are presented.

\subsection{Even Parity}

\noindent The evolution equation for the scalar is

\begin{equation}
    \dot{\phi}^S = \left(\Sigma-\frac{2\Theta}{3} \right)\left(\frac{1}{2}\phi^S-\mathcal{A}^S \right)-\frac{\okt}{a_2}\alpha^V,
\end{equation}

\bigskip

\noindent for the 2-vectors are

\begin{equation}
  \begin{aligned}
	\dot{\overline{\mathcal{H}}}^V &  = \frac{i\pk}{2a_1}\mathcal{E}^V+\frac{3}{4}\left(\Sigma-\frac{4\Theta}{3} \right)\overline{\mathcal{H}}^V-\frac{3}{4}X^V \\  & \quad- \frac{3\mathcal{E}}{2}\mathcal{A}^V+\frac{3\mathcal{E}}{4}a^V-\frac{\mathcal{R}a_2^2-\okt}{4a_2}\mathcal{E}^T,
    \end{aligned}
\end{equation}

\begin{equation}
    \dot{\overline{\Omega}}^V = -\left(\frac{\Sigma}{2}+\frac{2\Theta}{3} \right)\overline{\Omega}^V +\frac{1}{2a_2}\mathcal{A}^S-\frac{i\pk}{2a_1} \mathcal{A}^V,
\end{equation}

\begin{equation}
	\dot{\mu}^V=\frac{1}{2}\left(\Sigma-\frac{2\Theta}{3} \right)\mu^V-\Theta\left( \mu^V+p^V \right)-(\mu+p)W^V+\dot{\mu}\mathcal{A}^V,
\end{equation}

\begin{equation}
	\begin{aligned}
        \dot{X}^V & =2\left(\Sigma-\frac{2\Theta}{3} \right)X^V+\frac{3\mathcal{E}}{2}\left(V^V-\frac{2}{3}W^V \right) \\ 
        & \quad -\frac{1}{2}(\mu+p)V^V  -\frac{\Sigma}{2}\left( \mu^V+p^V \right)+\dot{\mathcal{E}}\mathcal{A}^V+\frac{\okt}{a_2^2}\overline{\mathcal{H}}^V,
    \end{aligned}
\end{equation}

\begin{equation}
	\begin{aligned}
        \dot{V}^V-\frac{2}{3}\dot{W}^V & =\frac{1}{3}\left( \mu^V+3p^V \right) + \left(\dot{\Sigma}-\frac{2\dot{\Theta}}{3} + \frac{\okt}{a_2^2} \right)\mathcal{A}^V \\
        & \quad +\frac{3}{2}\left(\Sigma-\frac{2\Theta}{3} \right)\left(V^V-\frac{2}{3}W^V \right) - X^V,
    \end{aligned}
\end{equation}

\begin{equation}
    \dot{a}^V=\frac{i\pk}{a_1}\alpha^V-\left(\Sigma+\frac{\Theta}{3} \right)\left( \mathcal{A}^V+a^V \right)-\overline{\mathcal{H}}^V,
\end{equation}

\begin{equation}
  \begin{aligned}
	\dot{W}^V & = \left(\dot{\Theta}-\frac{\okt}{a_2^2} \right)\mathcal{A}^V + \frac{i\pk}{a_1a_2}\mathcal{A}^S-\frac{1}{2}\left(\mu^V+3p^V \right) \\ 
    & \quad +\left(\frac{\Sigma}{2}-\Theta \right) W^V-3\Sigma V^V,
    \end{aligned}
\end{equation}

\begin{equation}
    \dot{\Sigma}^V = \frac{i\pk}{2a_1} \mathcal{A}^V +  \frac{1}{2a_2}\mathcal{A}^S - \left(\frac{\Sigma}{2}+\frac{2\Theta}{3} \right)\Sigma^V-\frac{3\Sigma}{2}\alpha^V-\mathcal{E}^V,
\end{equation}

\begin{equation}
	\begin{aligned}
        \dot{\mathcal{E}}^V & = \frac{i\pk}{2a_1} \overline{\mathcal{H}}^V+\frac{3}{4}\left(\Sigma - \frac{4\Theta}{3} \right)\mathcal{E}^V+\frac{1}{4}\left(3\mathcal{E}-2\mu-2p \right)\Sigma^V \\
        & \quad -\frac{3\mathcal{E}}{2}\alpha^V+\frac{\mathcal{R}a_2^2-\okt}{4a_2}\overline{\mathcal{H}}^T-\frac{3\mathcal{E}}{4}\Omega^V,
    \end{aligned}
\end{equation}

\noindent and for the 2-tensors are

\begin{equation}
    \dot{\overline{\mathcal{H}}}^T=-\frac{i\pk}{a_1}\mathcal{E}^T+\frac{1}{a_2}\mathcal{E}^V-\left(\Theta+\frac{3\Sigma}{2} \right)\overline{\mathcal{H}}^T+\frac{3\mathcal{E}}{2}\zeta^T,
\end{equation}

\begin{equation}
  \begin{aligned}
    \dot{\mathcal{E}}^T & = -\frac{i\pk}{a_1}\overline{\mathcal{H}}^T-\frac{1}{a_2}\overline{\mathcal{H}}^V-\frac{1}{2}(3\mathcal{E}+\mu+p)\Sigma^T \\ 
    & \quad -\left(\Theta+\frac{3\Sigma}{2} \right)\mathcal{E}^T,
    \end{aligned}
\end{equation}

\begin{equation}
	\dot{\zeta}^T=\frac{1}{2}\left(\Sigma-\frac{2\Theta}{3} \right)\zeta^T+\frac{1}{a_2}\alpha^V+\overline{\mathcal{H}}^T,
\end{equation}

\begin{equation}
	\dot{\Sigma}^T = \left(\Sigma-\frac{2\Theta}{3} \right)\Sigma^T+\frac{1}{a_2}\mathcal{A}^V-\mathcal{E}^T.
\end{equation}

\noindent The constraints for the scalars are

\begin{equation}
  \begin{aligned}
    \frac{i\pk}{a_1a_2}\phi^S & = \frac{1}{3}\left(\Sigma+\frac{4\Theta}{3} \right)W^V+\left(\frac{\Theta}{3}-2\Sigma \right)V^V \\ 
    & \quad-\frac{\okt}{a_2^2}a^V-\frac{2}{3}\mu^V-X^V,
    \end{aligned}
\end{equation}

\begin{equation}
    \ik p^V = -\frac{1}{a_2}\left(\mu+p \right)\mathcal{A}^S-2\dot{p}\overline{\Omega}^V,
\end{equation}

\noindent for the 2-vectors are

\begin{equation}
  \begin{aligned}
	\ik\left(V^V-\frac{2}{3}W^V \right) & = \frac{\okt}{a_2^2}\Sigma^V-\frac{3\Sigma}{2a_2}\phi^S \\ 
    & \quad -\frac{2}{3}\left(3\dot{\Sigma}-2\dot{\Theta} + \frac{3\okt}{2a_2^2}\right)\overline{\Omega}^V,
    \end{aligned}
\end{equation}

\begin{equation}
	\ik \left(\Sigma^V+\overline{\Omega}^V \right) = \frac{1}{2}V^V+\frac{2}{3}W^V-\rka\Sigma^T-\frac{3\Sigma}{2} a^V,
\end{equation}

\begin{equation}
    \ik\left(X^V-\frac{1}{3}\mu^V \right) = \frac{\okt}{a_2^2}\mathcal{E}^V-\frac{3\mathcal{E}}{2a_2}\phi^S-\frac{2}{3}\left(3\dot{\mathcal{E}}-\dot{\mu} \right)\overline{\Omega}^V,
\end{equation}

\begin{equation}
  \begin{aligned}
    \ik\mathcal{E}^V & =\frac{1}{2}X^V+\frac{1}{3}\mu^V-\frac{3\mathcal{E}}{2}a^V \\ 
    & \quad -\rka\mathcal{E}^T+\frac{3\Sigma}{2}\overline{\mathcal{H}}^V,
    \end{aligned}
\end{equation}

\begin{equation}
    V^V-\frac{2}{3}W^V=2\overline{\mathcal{H}}^V - \frac{\mathcal{R}a_2^2-\okt}{a_2}\Sigma^T,
\end{equation}

\begin{equation}
  \begin{aligned}
	\mathcal{E}^V & = -\frac{1}{2}\left(\Sigma-\frac{2\Theta}{3} \right)\left(\Sigma^V+\overline{\Omega}^V \right) \\
    & \quad + \rka\zeta^T - \frac{1}{2a_2}\phi^S,
    \end{aligned}
\end{equation}

\begin{equation}
    p^V=-\left(\mu+p \right)\mathcal{A}^V,
\end{equation}

\noindent and for the 2-tensors are

\begin{equation}
  \begin{aligned}
    \ik\overline{\mathcal{H}}^V & = \rka\overline{\mathcal{H}}^T - \left(\mu+p-\frac{3\mathcal{E}}{2} \right)\overline{\Omega}^V \\ 
    & \quad -\frac{3\mathcal{E}}{2}\Sigma^V+\frac{3\Sigma}{2}\mathcal{E}^V,
    \end{aligned}
\end{equation}

\begin{equation}
    \ik\zeta^T = \left(\Sigma+\frac{\Theta}{3} \right)\Sigma^T+\frac{1}{a_2}a^V-\mathcal{E}^T,
\end{equation}

\begin{equation}
    \ik\Sigma^T = \frac{1}{a_2}\Sigma^V-\frac{1}{a_2}\overline{\Omega}^V+\overline{\mathcal{H}}^T+\frac{3\Sigma}{2}\zeta^T.
\end{equation}

\subsection{Odd Parity}

\noindent The evolution equations for the scalars are

\begin{equation}
    2\dot{\xi}^S = \left(\Sigma-\frac{2\Theta}{3} \right)\xi^S+\frac{\okt}{a_2}\overline{\alpha}^V+\mathcal{H}^S,
\end{equation}

\begin{equation}
    \dot{\Omega}^S = \left(\Sigma-\frac{2\Theta}{3} \right)\Omega^S + \frac{\okt}{2a_2}\overline{\mathcal{A}}^V,
\end{equation}

\begin{equation}
    \dot{\mathcal{H}}^S=\frac{3}{2}\left(\Sigma-\frac{2\Theta}{3} \right)\mathcal{H}^S-\frac{\okt}{a_2}\overline{\mathcal{E}}^V-3\mathcal{E}\xi^S,
\end{equation}

\noindent for the 2-vectors are

\begin{equation}
    \dot{\Omega}^V = -\left(\frac{\Sigma}{2}+\frac{2\Theta}{3} \right)\Omega^V + \frac{i\pk}{2a_1} \overline{\mathcal{A}}^V,
\end{equation}

\begin{equation}
	\dot{\overline{\mu}}^V=\frac{1}{2}\left(\Sigma-\frac{2\Theta}{3} \right)\overline{\mu}^V-\Theta\left( \overline{\mu}^V+\overline{p}^V \right)-(\mu+p)\overline{W}^V+\dot{\mu}\overline{\mathcal{A}}^V,
\end{equation}

\begin{equation}
	\begin{aligned}
        \dot{\overline{X}}^V & =2\left(\Sigma-\frac{2\Theta}{3} \right)\overline{X}^V+\frac{3\mathcal{E}}{2}\left(\overline{V}^V-\frac{2}{3}\overline{W}^V \right)+\dot{\mathcal{E}}\overline{\mathcal{A}}^V \\
        & \quad -\frac{1}{2}(\mu+p)\overline{V}^V -\frac{\Sigma}{2} \left(\overline{\mu}^V+\overline{p}^V \right),
    \end{aligned}
\end{equation}

\begin{equation}
	\begin{aligned}
        \dot{\overline{V}}^V-\frac{2}{3}\dot{\overline{W}}^V &  =  \frac{1}{3}\left( \overline{\mu}^V+3\overline{p}^V \right)-\overline{X}^V + \left(\dot{\Sigma}-\frac{2\dot{\Theta}}{3} \right)\overline{\mathcal{A}}^V \\
        & \quad +\frac{3}{2}\left(\Sigma-\frac{2\Theta}{3} \right)\left(\overline{V}^V-\frac{2}{3}\overline{W}^V \right),
    \end{aligned}
\end{equation}

\begin{equation}
    \dot{\overline{a}}^V=\frac{i\pk}{a_1}\overline{\alpha}^V-\left(\Sigma+\frac{\Theta}{3} \right)\left( \overline{\mathcal{A}}^V+\overline{a}^V \right)+\mathcal{H}^V,
\end{equation}

\begin{equation}
	\dot{\overline{W}}^V = \dot{\Theta}\overline{\mathcal{A}}^V -\frac{1}{2}\left(\overline{\mu}^V+3\overline{p}^V \right)+\left(\frac{\Sigma}{2}-\Theta \right) \overline{W}^V-3\Sigma \overline{V}^V,
\end{equation}

\begin{equation}
    \dot{\overline{\Sigma}}^V = \frac{i\pk}{2a_1} \overline{\mathcal{A}}^V - \left(\frac{\Sigma}{2}+\frac{2\Theta}{3} \right)\overline{\Sigma}^V-\frac{3\Sigma}{2}\overline{\alpha}^V-\overline{\mathcal{E}}^V,
\end{equation}

\begin{equation}
	\begin{aligned}
        \dot{\overline{\mathcal{E}}}^V & = -\frac{i\pk}{2a_1} \mathcal{H}^V+\frac{3}{4}\left(\Sigma - \frac{4\Theta}{3} \right)\overline{\mathcal{E}}^V+\frac{3\mathcal{E}}{4}\overline{\Omega}^V \\
        & \quad + \frac{3}{4a_2}\mathcal{H}^S -\frac{3\mathcal{E}}{2}\overline{\alpha}^V+\frac{\mathcal{R}a_2^2-\okt}{4a_2}\mathcal{H}^T \\ 
        & \quad +\frac{1}{4}\left(3\mathcal{E}-2\mu-2p \right)\overline{\Sigma}^V,
    \end{aligned}
\end{equation}

\begin{equation}
	\begin{aligned}
        \dot{\mathcal{H}}^V & = -\frac{i\pk}{2a_1}\overline{\mathcal{E}}^V+\frac{3}{4}\left(\Sigma-\frac{4\Theta}{3} \right)\mathcal{H}^V+\frac{3}{4}\overline{X}^V \\
        & \quad + \frac{3\mathcal{E}}{2}\overline{\mathcal{A}}^V-\frac{3\mathcal{E}}{4}\overline{a}^V-\frac{\mathcal{R}a_2^2-\okt}{4a_2}\overline{\mathcal{E}}^T,
    \end{aligned}
\end{equation}

\noindent and for the 2-tensors are

\begin{equation}
    \dot{\mathcal{H}}^T=\frac{i\pk}{a_1}\overline{\mathcal{E}}^T+\frac{1}{a_2}\overline{\mathcal{E}}^V-\frac{3}{2}\left(\Sigma+\frac{2\Theta}{3} \right)\mathcal{H}^T-\frac{3\mathcal{E}}{2}\overline{\zeta}^T,
\end{equation}

\begin{equation}
  \begin{aligned}
    \dot{\overline{\mathcal{E}}}^T & = \frac{i\pk}{a_1}\mathcal{H}^T-\frac{1}{a_2}\mathcal{H}^V-\frac{1}{2}(3\mathcal{E}+\mu+p)\overline{\Sigma}^T \\ 
    & \quad -\frac{3}{2}\left(\Sigma+\frac{2\Theta}{3} \right)\overline{\mathcal{E}}^T,
    \end{aligned}
\end{equation}

\begin{equation}
	\dot{\overline{\zeta}}^T=\frac{1}{2}\left(\Sigma-\frac{2\Theta}{3} \right)\overline{\zeta}^T-\frac{1}{a_2}\overline{\alpha}^V-\mathcal{H}^T,
\end{equation}

\begin{equation}
	\dot{\overline{\Sigma}}^T = \left(\Sigma-\frac{2\Theta}{3} \right)\overline{\Sigma}^T-\frac{1}{a_2}\overline{\mathcal{A}}^V-\overline{\mathcal{E}}^T.
\end{equation}

\noindent The constraints for the scalars are

\begin{equation}
	\ik\xi^S = \left(\Sigma+\frac{\Theta}{3} \right)\Omega^S+\frac{\okt}{2a_2}\overline{a}^V,
\end{equation}

\begin{equation}
	\label{eqB44}
	\ik\Omega^S=\frac{\okt}{a_2}\Omega^V,
\end{equation}

\begin{equation}
	\ik\mathcal{H}^S=\frac{\okt}{a_2}\mathcal{H}^V-\left(\mu+p+3\mathcal{E} \right)\Omega^S,
\end{equation}

\noindent for the 2-vectors are

\begin{equation}
  \begin{aligned}
    \ik\mathcal{H}^V & =-\rka\mathcal{H}^T - \left(\mu+p-\frac{3\mathcal{E}}{2} \right)\Omega^V \\ 
    & \quad +\frac{3\mathcal{E}}{2}\overline{\Sigma}^V-\frac{3\Sigma}{2}\overline{\mathcal{E}}^V + \frac{1}{2a_2}\mathcal{H}^S,
    \end{aligned}
\end{equation}

\begin{equation}
    \frac{\okt}{a_2}\left(\Omega^V-\overline{\Sigma}^V \right)=3\Sigma\xi^S-\mathcal{H}^S,
\end{equation}

\begin{equation}
    \overline{X}^V+\frac{2}{3}\overline{\mu}^V = \frac{1}{3}\left(\Sigma+\frac{4\Theta}{3} \right)\overline{W}^V+\left(\frac{\Theta}{3}-2\Sigma \right)\overline{V}^V,
\end{equation}

\begin{equation}
	\ik\left(\overline{V}^V-\frac{2}{3}\overline{W}^V \right) = \frac{2}{3}\left(3\dot{\Sigma}-2\dot{\Theta} \right)\Omega^V,
\end{equation}

\begin{equation}
  \begin{aligned}
	\ik \left(\overline{\Sigma}^V-\Omega^V \right) & = \frac{1}{2}\overline{V}^V+\frac{2}{3}\overline{W}^V-\frac{1}{a_2}\Omega^S \\ 
    & \quad +\rka\overline{\Sigma}^T-\frac{3\Sigma}{2} \overline{a}^V,
    \end{aligned}
\end{equation}

\begin{equation}
    \ik\left(\overline{X}^V-\frac{1}{3}\overline{\mu}^V \right) = \frac{2}{3}\left(3\dot{\mathcal{E}}-\dot{\mu} \right)\Omega^V,
\end{equation}

\begin{equation}
  \begin{aligned}
    \ik\overline{\mathcal{E}}^V & =\frac{1}{2}\overline{X}^V+\frac{1}{3}\overline{\mu}^V-\frac{3\mathcal{E}}{2}\overline{a}^V \\ 
    & \quad +\rka\overline{\mathcal{E}}^T-\frac{3\Sigma}{2}\mathcal{H}^V,
    \end{aligned}
\end{equation}

\begin{equation}
    \overline{V}^V-\frac{2}{3}\overline{W}^V= \frac{\mathcal{R}a_2^2-\okt}{a_2}\overline{\Sigma}^T - 2\mathcal{H}^V-\frac{2}{a_2}\Omega^S,
\end{equation}

\begin{equation}
	 \overline{\mathcal{E}}^V= -\frac{1}{2}\left(\Sigma-\frac{2\Theta}{3} \right)\left(\overline{\Sigma}^V-\Omega^V \right) -\rka\overline{\zeta}^T + \frac{1}{a_2}\xi^S,
\end{equation}

\begin{equation}
	\label{eqB55}
    \overline{p}^V=-\left(\mu+p \right)\overline{\mathcal{A}}^V,
\end{equation}

\begin{equation}
    \ik \overline{p}^V =2\dot{p}\Omega^V,
\end{equation}

\noindent and for the 2-tensors are

\begin{equation}
    \ik\overline{\zeta}^T = \left(\Sigma+\frac{\Theta}{3} \right)\overline{\Sigma}^T-\frac{1}{a_2}\overline{a}^V-\overline{\mathcal{E}}^T,
\end{equation}

\begin{equation}
    \ik\overline{\Sigma}^T = -\frac{1}{a_2}\overline{\Sigma}^V-\frac{1}{a_2}\Omega^V-\mathcal{H}^T+\frac{3\Sigma}{2}\overline{\zeta}^T.
\end{equation}

\section{Harmonic Coefficients}
\label{HarmonicCoefficients}

From the equations in appendix \ref{LinearizedHarmonics} we can now solve for 27 of the harmonics coefficients in terms of the five even parity independent coefficients
\begin{equation}
    \label{evenfree}
	\left\{ \overline{\Omega}^V, \mu^V, \Sigma^T, \mathcal{E}^T, \overline{\mathcal{H}}^T \right\}
\end{equation}
and the three odd parity independent coefficients, 
\begin{equation}
    \label{oddfree}
	\left\{ \Omega^S,  \overline{\mathcal{E}}^T, \mathcal{H}^T \right\}
\end{equation}
respectively.

\subsection{Even parity}
The even harmonic coefficients can be expressed as

\begin{equation}
    p^V=c_s^2\mu^V,
\end{equation}

\begin{equation}
    \mathcal{A}^S=-\frac{a_2}{\mu+p}\left(2\dot{p}\overline{\Omega}^V+\frac{i\pk c_s^2}{a_1}\mu^V \right),
\end{equation}

\begin{equation}
    \mathcal{A}^V=-\frac{c_s^2}{\mu+p}\mu^V,
\end{equation}

\begin{equation}
	\frac{i\pk}{a_1}\zeta^T=\left(\Sigma+\frac{\Theta}{3} \right)\Sigma^T-\mathcal{E}^T,
\end{equation}

\begin{equation}
	\begin{aligned}
        \frac{ik_{\parallel }}{a_{1}a_{2}}\mathcal{E}^{V} & = \frac{2ik_{\parallel}}{a_1a_2B}\left(\mu + p \right)\left(\Sigma+\frac{\Theta}{3}\right)\overline{\Omega}^V + \frac{k_{\parallel }^{2}}{a_{1}^{2}a_{2}B}\mu^{V}  \\
        & \quad -\frac{3\mathcal{E}}{2}\left( \Sigma +\frac{\Theta }{3}\right) \Sigma ^{T}+\left( \frac{3\mathcal{E}}{2}-\frac{k_{\parallel }^{2}C}{a_{1}^{2}}\right) \mathcal{E}^{T}  \\ 
        & \quad -\frac{3ik_{\parallel }}{2a_{1}B}\left( 2\mathcal{E}\left( \Sigma +\frac{\Theta }{3}\right) \right. \\
        & \quad \left. +\frac{\Sigma}{a_{2}^{2}}\left({\mathcal{R}}a_2^2-k_{\perp }^{2}\right)\right)\overline{\mathcal{H}}^{T},
	\end{aligned}
\end{equation}

\begin{equation}
  \begin{aligned}
    \frac{ik_{\parallel }}{a_{1}a_{2}}\Sigma^{V} &  = \frac{ik_{\parallel }}{a_{1}a_{2}}\overline{\Omega}^V  -\frac{1}{2}\left( B+\frac{{\mathcal{R}}a_2^2-k_{\perp }^{2}}{a_{2}^{2}}\right)\Sigma^{T} \\ 
    & \quad +\frac{3\Sigma}{2} {\mathcal{E}}^{T}-\frac{ik_{\parallel }}{a_{1}}\overline{\mathcal{H}}^{T},
    \end{aligned}
\end{equation}

\begin{equation}
  \begin{aligned}
    \frac{2}{3a_{2}}\overline{\mathcal{H}}^{V} & = -\frac{2ia_1G}{3a_2k_{\parallel}B}\overline{\Omega}^V -\frac{\Sigma }{a_{2}B}\mu^{V}+\Sigma C{\mathcal{E}}^{T} \\
    & \quad -{\mathcal{E}}\Sigma ^{T}-\frac{ik_{\parallel }J}{3a_{1}}\overline{\mathcal{H}}^{T},
    \end{aligned}
\end{equation}

\begin{equation}
    \begin{aligned}
        \frac{W^{V}}{a_{2}} & = -\frac{a_1}{ik_{\parallel}a_2}\left(\frac{2k_{\parallel}^2}{a_1^2}+\frac{G}{B} \right)\overline{\Omega}^V +\frac{3\Sigma }{2}\left( 1-C\right) \mathcal{E}^{T}  \\ 
        & \quad +\frac{1}{2}\left( 3{\mathcal{E}}+\frac{{\mathcal{R}}a_2^2-k_{\perp }^{2}}{a_{2}^{2}}-B\right)\Sigma^{T} \\ 
        & \quad -\frac{ik_{\parallel }}{2a_{1}}\left(2-J\right) \overline{\mathcal{H}}^{T} + \frac{3\Sigma }{2a_{2}B}\mu^{V},
    \end{aligned}
\end{equation}

\begin{equation}
	\begin{aligned}
		\frac{V^{V}}{a_{2}} & = -\frac{4a_1}{3ia_2k_{\parallel}}\left(\frac{k_{\parallel}^2}{a_1^2}-\frac{G}{B} \right)\overline{\Omega}^V-\frac{B}{3}\left( 1+2C\right) \Sigma^{T} \\
        & \quad +\Sigma \left( 1+2C\right) \mathcal{E}^{T}-\frac{2ik_{\parallel }}{3a_{1}}\left( 1+J\right) \overline{\mathcal{H}}^{T}-\frac{2\Sigma }{a_{2}B}\mu^{V},
	\end{aligned}
\end{equation}

\begin{equation}
	\begin{aligned}
        \frac{ik_{\parallel }}{a_{1}a_{2}}\alpha^{V} & = -\frac{1}{a_{2}}\left( \frac{3\Sigma }{2B}+\left( \Sigma +\frac{\Theta }{3}\right) \frac{c_{s}^{2}}{\mu +p}\right)\mu^{V} \\ 
        & \quad -\frac{3\mathcal{E}}{2}\Sigma^{T}+\frac{3\Sigma C}{2}\mathcal{E}^{T}-\frac{ik_{\parallel }J}{2a_{1}}\overline{\mathcal{H}}^{T}+\frac{a_1G}{ia_2k_{\parallel}B}\overline{\Omega}^V,
	\end{aligned}
\end{equation}

\begin{equation}
	\begin{aligned}
		\frac{ik_{\parallel }}{a_{1}a_2^2}\phi^{S} & = -\frac{2ik_\parallel}{3a_1a_2\Sigma}\left(2\dot{\Sigma}-\frac{4\dot{\Theta}}{3}+\frac{2G}{B} \right)\overline{\Omega}^V-\frac{BL}{3 }\Sigma^{T} \\
        & \quad +\left( \Sigma L-\frac{k_{\perp }^{2}}{a_{2}^{4}B}\left(\mathcal{R}a_2^2-k_{\perp }^{2}\right)\right) \mathcal{E}^{T} \\
        & \quad -\frac{2ik_{\parallel }L}{3a_{1}}\overline{\mathcal{H}}^{T} -\frac{2k_{\parallel }^{2}}{a_{1}^{2}a_2B}\mu^{V},
	\end{aligned}
\end{equation}

\begin{equation}
    \begin{aligned}
        \frac{X^{V}}{a_{2}} & = \frac{4ik_{\parallel}}{3a_1a_2\Sigma}\left(\mu+p+\frac{G}{B}\left(1+\frac{9a_1^2\Sigma^2}{4k_{\parallel}^2} \right) \right)\overline{\Omega}^V  \\ 
        & \quad + \frac{1}{3a_{2}}\left(1-\frac{3}{B}\left( \frac{k_{\perp }^{2}}{a_{2}^{2}}+3\mathcal{E}\right)\right)\mu^{V}  \\ 
        & \quad + C\left( \frac{k_{\perp }^{2}}{a_{2}^{2}}+3\mathcal{E}\right) \mathcal{E}^{T}  +\frac{6a_{1}}{ik_{\parallel }B}\left( \frac{k_{\parallel }^{2}\mathcal{E}}{a_{1}^{2}}\left( \frac{\Theta}{3}-\frac{\Sigma}{2}\right) \right. \\ 
        & \quad \left. -\frac{k_{\perp }^{2}\Sigma}{4a_{2}^{4}}\left(\mathcal{R}a_2^2-k_{\perp }^{2}\right)\right) \overline{\mathcal{H}}^{T}-{\mathcal{E}}\left( \Theta-\frac{3\Sigma}{2}\right) \Sigma^{T},
    \end{aligned}
\end{equation}

where

\begin{equation}
    B\equiv \tilde{k}^2+\frac{9\Sigma ^{2}}{2}+3\mathcal{E},
\end{equation}
\begin{equation}
    CB\equiv \Sigma\left(\Theta-\frac{3\Sigma}{2}\right)-\frac{k_\perp^2}{a_2^2},
\end{equation}
\begin{equation}
	G\equiv\left(\mu+p \right)\left(\mathcal{R}-\tilde{k}^2 \right),
\end{equation}
\begin{equation}
    LB\equiv 3\Sigma\left(\frac{k_{\perp }^{2}}{a_{2}^{2}}-\frac{k_{\parallel }^{2}}{a_{1}^{2}}\right) +\Theta\tilde{k}^2,
\end{equation}
\begin{equation}
    JB\equiv \frac{k_\perp^2 a_1^2}{k_\parallel ^2 a_2^2}\left({\mathcal{R}}-\tilde{k}^2\right)+2\Sigma\left(\Theta-\frac{3\Sigma}{2}\right).
\end{equation}

\subsection{Odd parity}

For the odd harmonic coefficients the following equations hold,

\begin{equation}
	\Omega^V = \frac{ia_2k_\parallel}{a_1k_\perp^2}\Omega^S,
\end{equation}

\begin{equation}
	\overline{\mu}^V = \frac{2a_2\dot{\mu}}{k_\perp^2}\Omega^S,
\end{equation}

\begin{equation}
    \overline{V}^V=\frac{2a_2\dot{\Sigma}}{k_\perp^2}\Omega^S,
\end{equation}

\begin{equation}
    \overline{W}^V=\frac{2a_2\dot{\Theta}}{k_\perp^2}\Omega^S,
\end{equation}

\begin{equation}
    \overline{X}^V=\frac{2a_2\dot{\mathcal{E}}}{k_\perp^2}\Omega^S,
\end{equation}

\begin{equation}
    \overline{p}^V=\frac{2a_2\dot{p}}{k_\perp^2}\Omega^S,
\end{equation}

\begin{equation}
    \overline{\mathcal{A}}^V=-\frac{}{}\frac{2 a_2\dot{p}}{k_\perp^2 \left(\mu+p \right)}\Omega^S,
\end{equation}

\begin{equation}
    \frac{i\kp}{a_1}\xi^S=\left(\Sigma+\frac{\Theta}{3} \right)\Omega^S,
\end{equation}

\begin{equation}
  \begin{aligned}
    \frac{i a_2 k_\parallel B}{a_1 k_\perp^2}\mathcal{H}^S & =-\frac{2a_2}{k_\perp^2}(\mu+p)\left(B+\mathcal{R}-\frac{k_\perp^2}{a_2^2} \right)\Omega^S \\ 
    & \quad+\frac{1}{a_2}\left(\mathcal{R}a_2^2-k_\perp^2 \right)\left(\frac{3\Sigma}{2}\overline{\mathcal{E}}^T+\frac{ik_\parallel}{a_1}\mathcal{H}^T \right),
    \end{aligned}
\end{equation}

\begin{equation}
    \frac{B}{2}\overline{\Sigma}^T=-\frac{1}{k_\perp^2}\left(B+2\left( \mu+p \right) \right)\Omega^S+\frac{3\Sigma}{2}\overline{\mathcal{E}}^T+\frac{ik_\parallel}{a_1}\mathcal{H}^T,
\end{equation}

\begin{equation}
  \begin{aligned}
    \mathcal{H}^V & =\frac{1}{a_2B}\left(\mathcal{R}a_2^2-k_\perp^2 \right)\left(\frac{3\Sigma}{2}\overline{\mathcal{E}}^T+\frac{ik_\parallel}{a_1}\mathcal{H}^T \right) \\ 
    & \quad \frac{a_2}{k_\perp^2}\left(3\mathcal{E}-N \right)\Omega^S,
    \end{aligned}
\end{equation}

\begin{equation}
  \begin{aligned}
    \frac{ik_\parallel}{a_1}\overline{\alpha}^V & =-\frac{a_2}{k_\perp^2}\left(3\mathcal{E}-N -2\Theta c_s^2\left(\Sigma+\frac{\Theta}{3} \right) \right)\Omega^S \\ 
    & \quad - \frac{1}{a_2B}\left(\mathcal{R}a_2^2-k_\perp^2 \right)\left(\frac{3\Sigma}{2}\overline{\mathcal{E}}^T+\frac{ik_\parallel}{a_1}\mathcal{H}^T \right),
    \end{aligned}
\end{equation}

\begin{equation}
    \begin{aligned}
        3\Sigma\overline{\mathcal{E}}^V & = \frac{3\Sigma}{2a_2}\left(\mathcal{R}a_2^2-k_\perp^2 \right)\left( \frac{a_{1}}{ik_{\parallel }}\left( 1-\frac{9\Sigma ^{2}}{2B}\right)\overline{\mathcal{E}}^{T}-\frac{3\Sigma}{B}\mathcal{H}^{T}\right) \\ 
        & \quad +  \frac{ia_1a_2}{k_\parallel k_\perp^2}\left(\left(\mu+p+N \right)\left(\tilde{k}^2+3\mathcal{E} \right)\vphantom{\frac{k_{\parallel}^ 2}{a_1^ 2}} \right. \\
          & \quad \left. + 9\mathcal{E}\Sigma\left(\Sigma+\frac{\Theta}{3} \right)-\frac{4k_\parallel^2}{a_1^2}\left(\mu+p \right) \right)\Omega^S,
    \end{aligned}
\end{equation}

\begin{equation}
  \begin{aligned}
    \frac{ik_\parallel}{a_1}\overline{\Sigma}^V & =-\frac{a_2}{k_\perp^2}\left(3\Sigma^2+\Sigma\Theta+\mu+p+N+\frac{k_\parallel^2}{a_1^2} \right)\Omega^S \\ 
    & \quad +\frac{1}{a_2B}\left(\mathcal{R}a_2^2-k_\perp^2 \right)\left(\frac{3\Sigma}{2}\overline{\mathcal{E}}^T+\frac{ik_\parallel}{a_1}\mathcal{H}^T \right),
    \end{aligned}
\end{equation}

\begin{equation}
  \begin{aligned}
    \frac{B}{2}\overline{\zeta}^T & =-\frac{a_1}{ik_\parallel k_\perp^2}\left( B+2\left(\mu+p \right)\right)\left(\Sigma + \frac{\Theta}{3}\right)\Omega^S \\ 
    & \quad + \left( \Sigma +\frac{\Theta }{3}\right){\mathcal{H}}^{T}-\frac{a_{1}}{2ik_{\parallel}}\left(\tilde{k}^2-\mathcal{R}\right) \overline{\mathcal{E}}^{T}
    \end{aligned}
\end{equation}

with

\begin{equation}
    N\equiv \left(\mu  + p \right)\left(1+\frac{2}{a_2^2B}\left(\mathcal{R}a_2^2-k_\perp^2 \right) \right).
\end{equation}

%=====================================
% References, variant A: internal bibliography
%=====================================

\end{document}